\documentclass[twocolumn,showpacs,preprintnumbers,amsmath,amssymb]{revtex4}
\usepackage{epsfig}
\usepackage{dcolumn}
\usepackage{bm}

\newcommand{\remove}[1]{}
\newcommand{\dd}{\mathrm{d}}

\def\be{\begin{equation}}
\def\ee{\end{equation}}

\newcommand{\beq}{\begin{equation}}
\newcommand{\eeq}{\end{equation}}
\newcommand{\beqa}{\begin{eqnarray}}
\newcommand{\eeqa}{\end{eqnarray}}

\renewcommand{\pl}{\partial}
\newcommand{\lag}{\langle}
\newcommand{\rag}{\rangle}

\newcommand{\ii}{{\rm i}}

\newcommand{\vu}{{\bf u}}
\newcommand{\vv}{{\bf v}}

\newcommand{\vx}{{\bf x}}
\newcommand{\vk}{{\bf k}}

\newcommand{\vq}{{\bf q}}
\renewcommand{\vr}{{\bf r}}

\newcommand{\vOm}{{\bf \Omega}}

\newcommand{\tC}{{\tilde{C}}}
\newcommand{\tdelta}{{\tilde{\delta}}}

\newcommand{\tW}{{\tilde{W}}}

\newcommand{\cD}{{\cal D}}

\newcommand{\cG}{{\cal G}}

\newcommand{\cL}{{\cal L}}

\newcommand{\cbD}{{\textbf {\textsf D}}}

\newcommand{\rhob}{\overline{\rho}}
\newcommand{\Om}{\Omega_{\rm m}}

\newcommand{\bea}{\begin{array}}
\newcommand{\ea}{\end{array}}

\begin{document}

\title{Angular averaged consistency relations of large-scale structures}

\author{Patrick Valageas}
\affiliation{Institut de Physique Th\'eorique,\\
CEA, IPhT, F-91191 Gif-sur-Yvette, C\'edex, France\\
CNRS, URA 2306, F-91191 Gif-sur-Yvette, C\'edex, France}
\vspace{.2 cm}

\date{\today}
\vspace{.2 cm}

\begin{abstract}
The cosmological dynamics of gravitational clustering satisfies an approximate
invariance with respect to the cosmological parameters that is often used
to simplify analytical computations. We describe how this approximate symmetry
gives rise to angular averaged consistency relations for the matter density correlations.
This allows one to write the $(\ell+n)$ density correlation, with $\ell$
large-scale linear wave numbers that are integrated over angles, and $n$ fixed
small-scale nonlinear wave numbers, in terms of the small-scale $n$-point density correlation
and $\ell$ prefactors that involve the linear power spectra at the large-scale
wave numbers.
These relations, which do not vanish for equal-time statistics, go beyond the
already known kinematic consistency relations.
They could be used to
detect primordial non-Gaussianities, modifications of gravity, limitations
of galaxy biasing schemes, or to help designing analytical models of
gravitational clustering.

\end{abstract}

 \vskip2pc

\maketitle

\section{Introduction}
\label{Introduction}

After the results of the WMAP and Planck missions 
\cite{Komatsu2011,PlanckXVI-2013}, which have already uncovered
a lot of information from the cosmic microwave background data, 
surveys of the large-scale structure of the Universe promise to be an important
and complementary probe of cosmological scenarios
\cite{Albrecht2006,Laureijs2011}.
In particular, they should shed light on the properties of the 
dark matter and dark energy components. 
Unfortunately, even without considering the very
complex processes of galaxy and star formation 
\cite{Scannapieco2012,Bryan2013,Semboloni2013,Martizzi2013}
and focusing on the large-scale
properties where gravity is the dominant driver, exact or well-controled
predictions for the statistical properties of the density and velocity fields are
difficult.
Large scales can be described by standard perturbative approaches
\cite{Goroff1986,Bernardeau2002},
which can be improved to some degree by using resummation schemes
\cite{Crocce2006a,Valageas2007,Pietroni2008,Bernardeau2008,Taruya2012,Pietroni2012,Crocce2012,Bernardeau2013,Valageas2013}.
However, these methods cannot reach the truly nonlinear regime where
shell-crossing effects become important 
\cite{Pueblas2009,Valageas2011a,Valageas2013a}.
Small scales are studied through numerical
simulations or phenomenological models \cite{Cooray2002}
that rely on informations gained through these simulations.
However, these scales are difficult to model with a high accuracy, even with 
simulations, and it would be useful to have analytical results that go beyond
low-order perturbation theory.

Some exact results have recently been obtained 
\cite{Kehagias2013,Peloso2013,Creminelli2013,Kehagias2013a,Peloso2013a,Creminelli2013a,Valageas2013b}
in the form of ``kinematic consistency relations''. 
They relate the $(\ell+n)$-density correlation, with $\ell$ large-scale wave numbers
and $n$ small-scale wave numbers, to the $n$-point small-scale density correlation,
with $\ell$ prefactors that involve the linear power spectrum at the
large-scale wave numbers.
These relations, obtained at the leading order over the large-scale wave numbers $k_j'$,
arise from the equivalence principle (in standard scenarios). It ensures
that small-scale structures respond to a large-scale perturbation (which at leading
order corresponds to a constant gravitational force over the extent of the small-size
object) by a uniform displacement. Therefore, these relations express
a kinematic effect, due to the displacement of small-scale structures between
different times.
This also means that (at this order) they vanish for equal-time statistics, as
a uniform displacement has no impact on the statistical properties of the density
field observed at a given time.

In practice, it is difficult to measure different-time density correlations,
as correlations between different redshift planes along our light cone 
(hence over distances of order $c/H_0$) are very small.
Therefore, it would be useful to obtain similar relations that apply to single-time
density correlations. This means that we must go beyond the kinematic effect
and investigate how small-scale density fluctuations respond to non-uniform
gravitational forces. At leading order over the large-scale wave numbers, 
this is given by the response to a change of the background density, which also
corresponds to a large-scale curvature of the gravitational potential.

In this paper, we show how this problem can be addressed by using an
approximate symmetry of the cosmological gravitational dynamics.

In Sec.~\ref{approx-sym}, we recall how most of the dependence on cosmological
parameters can be absorbed by a remapping of the time-coordinate,
$t \rightarrow D_+(t)$, where $D_+(t)$ is the linear growing mode.
This is a well-know approximate symmetry of the cosmological gravitational dynamics
that is often used in analytical methods (e.g., perturbative schemes) to simplify
the computations.
Then, in Sec.~\ref{consist} we show how this invariance dictates the response
of density fluctuations to a small change of the background density,
which corresponds to a change of the cosmological parameters $\Om$ and
$\Omega_K$. This allows us to derive consistency relations that go beyond the
kinematic effect and remain nontrivial for the
equal-time density correlations. 
In Sec.~\ref{checks}, we explicitly check this relation for the
matter density bispectrum, at leading order of perturbation theory.
We also present a fully nonlinear check in 1D (using the fact that the Zel'dovich
approximation becomes an exact solution), which provides a check for all
many-body density correlations or polyspectra up to all orders of perturbation
theory (and beyond the shell crossing regime if we consider the system as defined
by the Zel'dovich solution).
We discuss our results and conclude in Sec.~\ref{conclusion}.

\section{Approximate symmetry of the cosmological gravitational dynamics}
\label{approx-sym}

On scales much smaller than the horizon, where the Newtonian approximation
is valid, the equations of motion read as \cite{Peebles1980}
\beq
\frac{\pl\delta}{\pl t} + \frac{1}{a} \nabla \cdot [ (1+\delta) \vv ] = 0 , 
\label{cont-1}
\eeq
\beq
\frac{\pl\vv}{\pl t} + H \vv + \frac{1}{a} (\vv\cdot\nabla)\vv = 
- \frac{1}{a} \nabla \phi ,
\label{Euler-1}
\eeq
\beq
\nabla^2 \phi = 4\pi \cG \rhob a^2 \delta , 
\label{Poisson-1}
\eeq
where $a(t)$ is the scale factor, $H=\dot{a}/a$ the Hubble expansion rate, 
$\delta=(\rho-\rhob)/\rhob$ the density contrast, and $\vv=a\dd\vx/\dd t$ the peculiar
velocity. Here, we use the single-stream approximation to simplify the presentation,
but the results remain valid beyond shell crossing.
Linearizing these equations over $\{\delta,\vv\}$, one obtains the linear growth rates
$D_\pm(t)$, which are the independent solutions of 
\cite{Peebles1980,Bernardeau2002}
\beq
\ddot{D} + 2 H \dot{D} - 4\pi\cG\rhob D = 0 .
\label{Dlin-1}
\eeq
For an Einstein-de Sitter universe, where $a(t)\propto t^{2/3}$, the linear growing mode
is $D_+(t) \propto a \propto t^{2/3}$ and the linear decaying mode is 
$D_-(t) \propto a^{-3/2} \propto t^{-1}$.
For a generic cosmology, with a nonzero cosmological constant and curvature,
one must numerically solve Eq.(\ref{Dlin-1}).

As usual \cite{Crocce2006a,Crocce2006,Valageas2008}, it is convenient to make the change of variables
\beq
\eta = \ln D_+ , \;\;\; \vv = \dot{a} f \vu , \;\;\; \phi = (\dot{a} f)^2 \varphi ,
\label{eta-3D}
\eeq
where $f=\dd\ln D_+/\dd\ln a$. Then, the equations of motion read as
\beq
\frac{\pl\delta}{\pl\eta} + \nabla \cdot [ (1+\delta) \vu ] = 0 , 
\label{cont-2}
\eeq
\beq
\frac{\pl\vu}{\pl \eta} + \left( \frac{3\Om}{2f^2} - 1 \right) \vu + (\vu\cdot\nabla)\vu 
= - \nabla \varphi ,
\label{Euler-2}
\eeq
\beq
\nabla^2 \varphi = \frac{3\Om}{2f^2} \, \delta , 
\label{Poisson-2}
\eeq
where $\Om(t)$ is the matter density cosmological parameter as a function of time,
which obeys $4\pi\cG\rhob=(3/2) \Om H^2$.

It happens that for standard cosmologies (i.e., within General Relativity), $\Om/f^2$
is always very close to $1$ (which is exact for the Einstein-de Sitter case)
\cite{Peebles1980}.
Then, making the approximation $\Om/f^2 \simeq 1$ removes all explicit time
dependence in the equations of motion (\ref{cont-2})-(\ref{Poisson-2}) and 
simplifies the analytical computations. This also removes all explicit dependence
on the cosmological parameters.
In particular, within a perturbative framework, one can use the results obtained for 
the Einstein-de Sitter case by making the replacement $a(t) \rightarrow D_+(t)$
\cite{Nusser1998,Scoccimarro1998}.
The accuracy of this approximation was investigated by 
Refs.\cite{Pietroni2008,Crocce2012},
who find that it performs to better than $1\%$ at redshift $z=0$ for
$k \leq 2h$Mpc$^{-1}$, and $0.1\%$ at $z=1$ on these scales.
The approximation performs increasingly well at high redshift in the matter era 
(where we recover the Einstein-de Sitter cosmology). Although it degrades
on small scales at $z=0$, this approximation is used by most perturbative approaches
\cite{Crocce2006a,Valageas2007,Pietroni2008,Bernardeau2008,Taruya2012,Pietroni2012,Crocce2012,Bernardeau2013,Valageas2013}
to simplify computations (in particular, it allows one to use the explicit exponential
form of the linear response function or propagator adapted from the
Einstein - de Sitter case, with factors $e^{\eta}$ and $e^{-3\eta/2}$
\cite{Crocce2006a,Crocce2006b,Valageas2007}).
Thus, it provides a sufficient approximation on perturbative scales and in the
highly nonlinear regime at low redshift it is not the main source of inaccuracy,
as uncertainties on the halo mass function for instance lead to greater error bars
\cite{Valageas2013a}.

This approximate symmetry does not rely on the single-stream approximation, 
and instead of the Euler equations (\ref{Euler-1}) and (\ref{Euler-2}), we can use the 
equation of motion of the trajectories $\vx(\vq,t)$ of the particles. It reads as
\beq
\frac{\pl^2\vx}{\pl t^2} +  2 H \frac{\pl\vx}{\pl t} = - \frac{1}{a^2} \nabla \phi ,
\label{traj-1}
\eeq
which becomes with the time coordinate $\eta$
\beq
\frac{\pl^2\vx}{\pl\eta^2} +  \left( \frac{3\Om}{2f^2} - 1 \right) 
\frac{\pl\vx}{\pl\eta} = - \nabla \varphi ,
\label{traj-2}
\eeq
where $\varphi$ is the rescaled gravitational potential (\ref{Poisson-2}).
This explicitly shows that it satisfies the same approximate symmetry.
Therefore, our results are not restricted to the perturbative regime
and also apply to small nonlinear scales governed by shell-crossing effects,
as long as the approximation $\Om/f^2 \simeq 1$ is sufficiently accurate
(but this also means that we are restricted to scales dominated by gravity).

This standard approximation means that all the dependence on cosmological
parameters is encapsulated in the linear growing mode $D_+(t)$.
In this paper, we investigate the consequences of this approximate symmetry of the
equations of motion, in terms of the ``squeezed'' limit of density correlations.
This corresponds to Fourier space $(\ell+n)$ density correlations 
$\lag \tdelta(\vk_1') .. \tdelta(\vk_{\ell}') \tdelta(\vk_1) .. \tdelta(\vk_n) \rag$, where the 
wave numbers $k_j'$
are much smaller than all other wavenumbers $k_i$ and within the linear regime.
Our method relies on the fact that a large-scale spherically symmetric perturbation of the
initial density contrast is similar to a change of the mean density $\rhob$, whence
of the cosmological parameters, from the point of view of a much smaller region at the
center of this initial perturbation.

\section{Angular averaged consistency relations}
\label{consist}

We first consider the case of a single large-scale wave number $k'$ and we
generalize to several large-scale wave numbers $k_j'$
in Sec.~\ref{several}.

\subsection{Correlation and response functions}
\label{Corr-resp}

Because the cosmological density and velocity fields are statistically homogeneous
and isotropic, it is often convenient to work in Fourier space. In this paper, we denote
with a tilde Fourier-space fields, defining the Fourier transform as
\beq
\delta(\vx) = \int \dd \vk \; e^{\ii\vk\cdot\vx} \; \tdelta(\vk) .
\label{Fourier-def}
\eeq
To compare theoretical predictions with observations, one often computes
correlation functions, $\lag \delta(\vx_1) \dots \delta(\vx_n) \rag$,
or multispectra, $\lag \tdelta(\vk_1) \dots \tdelta(\vk_n)\rag$.
In particular, the power spectrum $P(k)$ is defined as
\beq
\lag \tdelta(\vk_1) \tdelta(\vk_2) \rag = \delta_D(\vk_1+\vk_2) \; P(k_1) ,
\label{Pk-def}
\eeq
where the Dirac factor arises from statistical homogeneity.
We also denote with the subscript ``L'' the linear fields obtained by linearizing
the equations of motion (\ref{cont-1})-(\ref{Poisson-1}), and with the subscript
``L0'' the linear fields today, at $z=0$.
Throughout this paper, we assume as usual that the linear decaying modes have
had time to become negligible by the times of interest. Then, the initial conditions are
fully defined by the linear growing mode, which is also set by the linear density
field today $\delta_{L0}(\vx)$, which we assume to be Gaussian.

In analytical approaches, especially in perturbative schemes that use
field-theoretic tools 
\cite{Crocce2006a,Crocce2006b,Valageas2007,Valageas2007a,Taruya2008,Bernardeau2008,Anselmi2011,Bernardeau2012a}, it is convenient to introduce response functions 
(also called propagators or Green functions), that we define in real space as
\beqa
R^{\ell,n}(\vx_1',..,\vx_{\ell}' ; \vx_1,t_1,..,\vx_n,t_n) & = & \nonumber \\
&& \hspace{-2.3cm} \left \lag \frac{ 
\cD^{\ell}[ \delta(\vx_1,t_1) .. \delta(\vx_n,t_n) ]}
{\cD \delta_{L0}(\vx_1') .. \cD\delta_{L0}(\vx_{\ell}')} \right \rag , \;\;\;
\label{Resp-def}
\eeqa
and similarly in Fourier space (throughout this paper, we denote by the letter
$\cD$ the functional derivative).
These quantities (\ref{Resp-def}) describe how the nonlinear 
density field, at positions and times $\{\vx_1,t_1;..;\vx_n,t_n\}$, responds to
changes of the initial conditions [defined by $\delta_{L0}(\vx)$] at positions
$\{\vx_1',..,\vx_{\ell}'\}$.

As described for instance in \cite{Valageas2013b}, for Gaussian initial conditions,
correlations between the nonlinear density
contrast $\delta$ and the linear density contrast $\delta_{L0}$ that
defines the initial conditions can be written in terms of 
response functions. This gives in Fourier space
\cite{Valageas2013b}
\beqa
\lag \tdelta_{L0}(\vk') \tdelta(\vk_1,t_1) .. \tdelta(\vk_n,t_n) \rag  & = & P_{L0}(k') 
\nonumber \\
&& \hspace{-1.8cm} \times \left \lag \frac{\cD[\tdelta(\vk_1,t_1) .. \tdelta(\vk_n,t_n)]}
{\cD\tdelta_{L0}(-\vk')} \right \rag  , \;\;\;
\label{Cn-Rn-1}
\eeqa
where $P_{L0}(k')$ is the linear power spectrum of the initial conditions $\delta_{L0}$.
This provides a simple method to obtain consistency relations for the density
correlations by computing the response function [i.e., the last term in 
Eq.(\ref{Cn-Rn-1})] associated with a large-scale perturbation $\Delta \delta_{L0}$
 of the initial condition.

The leading-order effect that arises in the large-scale limit, $k'\rightarrow 0$,
 is a constant force, 
$-\nabla( \Delta \phi_{L0})$, and velocity, $\Delta \vv_{L0}$, over the small-scale region
of size $R$, with $(k'R) \ll 1$.
This also corresponds to a zero local density perturbation, because in the linear regime
we have $\delta_L \propto \nabla\vv_L$ (up to time-dependent factors), as seen
from the continuity equation (\ref{cont-1}).
This leads to a uniform displacement of small-scale structures.
Then, one obtains kinematic consistency relations
\cite{Kehagias2013,Peloso2013,Creminelli2013,Peloso2013a,Creminelli2013a,Kehagias2013a,Valageas2013b}
that express a correlation of the form  $\lag \tdelta_{L0}(\vk_1') .. \tdelta_{L0}(\vk_{\ell}') 
\tdelta(\vk_1,t_1) .. \tdelta(\vk_n,t_n) \rag$, with $\ell$ low wave numbers and $n$ high
wave numbers, as a product of $\ell$ linear power spectra $P_{L0}(k_j')$
with the small-scale nonlinear correlation $\lag \tdelta(\vk_1,t_1) .. \tdelta(\vk_n,t_n) \rag$,
at lowest order over $k_j'$.
Because this corresponds to a uniform displacement, this leading-order result
vanishes at equal times, $t_1=..=t_n=t$, and the results obtained for
different times $t_1 \neq .. \neq t_n$ simply describe how small-scale patches
have moved in-between these times because of the force exerted by a large-scale
perturbation.

In this paper, we go beyond the kinematic effect recalled above and
we consider the effect of a nonzero large-scale density fluctuation, that is,
a nonzero curvature of the gravitational potential.
This higher-order effect does not vanish for equal-time statistics because the
large-scale perturbation of the gravitational potential curvature leads to a
deformation of the small-scale structure (mostly a space-time dilatation, as the
overall collapse is accelerated or decelerated).
This leads to consistency relations for density correlations that remain
nontrivial for single-time correlations.
To remove constant gradients, which are absorbed by the kinematic effect and do
not contribute to equal-time statistics, and to mimic a constant large-scale density
fluctuation (and isotropic curvature of the gravitational potential), we consider 
spherical averages that write in configuration space as
\beq
C^n_W = \int \dd\vx' W(x') \, \lag \delta_{L0}(\vx') \delta(\vx_1,t_1) .. \delta(\vx_n,t_n) \rag ,
\label{Cn-Wx-1}
\eeq
and in Fourier space as
\beq
\tC^n_W = (2\pi)^3 \int \dd\vk' \tW(k') \, \lag \tdelta_{L0}(\vk') \tdelta(\vk_1,t_1) .. 
\delta(\vk_n,t_n) \rag ,
\label{Cn-Wk-1}
\eeq
where $W(x')$, and its Fourier transform $\tW(k')$, is a large-scale window function.
Using Eq.(\ref{Cn-Rn-1}) and its configuration-space counterpart, 
Eqs.(\ref{Cn-Wx-1}) and (\ref{Cn-Wk-1}) read as
\beq
C^n_W = \!\! \int \!\! \dd\vx\dd\vx' W(x) \, C_{L0}(\vx,\vx') \left \lag 
\frac{\cD[ \delta(\vx_1,t_1) .. \delta(\vx_n,t_n) ]}{\cD\delta_{L0}(\vx')} \right \rag ,
\label{Cn-Wx-2}
\eeq
where $C_{L0}$ is the linear density correlation of the initial conditions, and
\beq
\tC^n_W = (2\pi)^3 \!\! \int \!\! \dd\vk' \tW(k') \, P_{L0}(k') \left \lag \frac{\cD[ \tdelta(\vk_1,t_1) .. 
\tdelta(\vk_n,t_n) ]}{\cD\tdelta_{L0}(-\vk')} \right \rag .
\label{Cn-Wk-2}
\eeq

By definition of the functional derivatives, these expressions also mean that we must
consider the change of the small-scale density correlation at linear order over
a perturbation $\Delta\delta_{L0}$, as
\beq
C^n_W = \left. \frac{\dd}{\dd\varepsilon} \right |_{\varepsilon=0} 
\lag \delta(\vx_1,t_1) .. \delta(\vx_n,t_n) \rag_{\varepsilon} 
\label{Cnx-eps-1}
\eeq
and
\beq
\tC^n_W = \left. \frac{\dd}{\dd\varepsilon} \right |_{\varepsilon=0} 
\lag \tdelta(\vk_1,t_1) .. \tdelta(\vk_n,t_n) \rag_{\varepsilon} ,
\label{Cnk-eps-1}
\eeq
where $\lag .. \rag_{\varepsilon}$ is the statistical average with respect to the Gaussian
initial conditions $\delta_{L0}$, when the linear density field is modified as
\beq
\delta_L(\vx) \rightarrow \delta_L(\vx) + \varepsilon D_+(t) \int \dd\vx' W(x') 
C_{L0}(\vx,\vx') 
\label{eps-x-1}
\eeq
or
\beq
\tdelta_L(\vk) \rightarrow \tdelta_L(\vk) + \varepsilon D_+(t) (2\pi)^3 \tW(k) P_{L0}(k) .
\label{eps-k-1}
\eeq
Here and in the following, we normalize the linear growth rate and the linear
density field as $\delta_L(\vx,t) = D_+(t) \delta_{L0}(\vx)$.
The spherical average over a much larger scale than the region of interest
of size $R$ in Eq.(\ref{eps-x-1}) ensures that over this small patch the density
perturbation $\Delta \delta_{L0} \simeq \varepsilon \int\dd\vx' W(x') C_{L0}(x')$
is constant (at leading order over $k'R$).

A similar idea was investigated in Ref.~\cite{Creminelli2013b} in the context
of single-field inflation, noticing that the effect of a large-scale fluctuation is
similar to changing the curvature of the universe, from the point of view of a 
small-scale region. However, this leads to a consistency relation between
a $(1+n)$ correlation such as Eq.(\ref{Cn-Wx-1}) and a small-scale $n$-point
correlation in a different universe. As such, it cannot be directly measured
because we have access to only one universe (unless one compares different
large-scale regions characterized by different large-scale mean densities).
In this paper, focusing on the late-time universe during the matter and dark
energy epochs, we show in the next section how the approximate symmetry
recalled in Sec.~\ref{approx-sym} allows us to derive consistency relations between 
correlations measured in the same universe. This is because this symmetry
provides a link between the cosmological gravitational dynamics in different
Friedmann-Lemaitre-Robertson-Walker cosmologies.

\subsection{Effect of a large-scale density perturbation}
\label{large-scale}

From the point of view of a small region, a much larger-scale almost uniform
perturbation to the initial density contrast is similar to a change of the background
density $\rhob$.
Then, following \cite{Peebles1980}, we first recall that such a small change
of the background also corresponds to a linear growing mode of the density
contrast.
Thus, we consider two universes with close cosmological parameters,
defined at the background level by the functions
$\{\rhob(t),a(t)\}$ and $\{\rhob'(t),a'(t)\}$.
The dynamics of the reference universe (i.e., our Universe) is given by the
Friedmann equations,
\beq
\frac{\dot{a}^2}{a^2} = \frac{8\pi\cG}{3} ( \rhob+\rhob_{\Lambda} ) 
- \frac{K}{a^2} ,
\label{Fried-1}
\eeq
\beq
\frac{\ddot{a}}{a} = - \frac{4\pi\cG}{3} \rhob + \frac{8\pi\cG}{3} \rhob_{\Lambda} ,
\label{Fried-2}
\eeq
where we included the contributions from a cosmological constant and a curvature
term and the dot denotes a derivative with respect to the time $t$.
The auxiliary universe $\{\rhob'(t),a'(t)\}$, denoted with a prime, obeys the same 
equations with the change $\{\rhob,a,K\} \rightarrow \{\rhob',a',K'\}$ (the constant
dark energy density $\rhob_{\Lambda}$ is not changed).
It only differs from the reference universe by a small amount, of order $\epsilon$,
with
\beq
\rhob a^3 = \rhob' a'^3 = \rhob_0 , \;\;\;
a' = a [1-\epsilon(t) ] , \;\;\;
\rhob' = \rhob [ 1+3\epsilon(t) ] .
\label{eps-def}
\eeq
Here and in the following, we only keep terms up to linear order over $\epsilon$.
Substituting Eq.(\ref{eps-def}) into the second Friedmann equation (\ref{Fried-2})
written for the auxiliary universe, we obtain 
\beq
\ddot{\epsilon} + 2 H \, \dot{\epsilon} - 4\pi\cG\rhob \, \epsilon =0 .
\label{eps-t}
\eeq
As is well known \cite{Peebles1980}, we recover the evolution equation 
(\ref{Dlin-1}) of the linear growth rates $D_{\pm}(t)$.  
This is because spherically symmetric shells evolve independently as separate
universes (before shell crossing), thanks to Birkhoff's theorem, and their
density difference behaves as the linear growth rate (in the linear regime).
Thus, we write
\beq
\epsilon(t) =  D_+(t) \, \epsilon_0 .
\label{eps0-def}
\eeq

We now turn to the density and velocity fluctuations.
In the reference universe, they follow the
equations of motion (\ref{cont-1})-(\ref{Poisson-1}). In the auxiliary universe, we have
the same equations of motion with primed variables. 
For our purpose, these two sets of variables actually describe the same physical
system, with two different choices for the background density $\rhob$ around which
we study fluctuations.
For instance, in the case $\epsilon > 0$, a density contrast $\delta'$ with a zero mean
in the primed frame appears as a density contrast $\delta$ with a nonzero positive
mean. Thus, a large-scale uniform density fluctuation $\Delta \delta_{L0}$
in the reference frame is absorbed by going to the primed frame. This will allow us
to study the effect of a large scale density fluctuation as in 
Eqs.(\ref{eps-x-1})-(\ref{eps-k-1}).

Because both frames refer to the same physical system, we have
\beq
\vr' = \vr = a' \vx' = a \vx , \;\;\;  \rhob' (1+\delta') = \rhob (1+\delta) ,
\eeq
where $\vr=\vr'$ is the physical coordinate. Thus, we have the relations
\beq
\vx' = (1+\epsilon) \vx , \;\;\; 
\delta' = \delta - 3 \epsilon (1+\delta) , \;\;\;
\vv' = \vv + \dot{\epsilon} a \vx , 
\label{x-d-v-1}
\eeq
where we used Eq.(\ref{eps-def}) and only kept terms up to linear order over 
$\epsilon$.
Then, we can check that if the fields $\{\delta',\vv',\phi'\}$ satisfy the equations of 
motion (\ref{cont-1})-(\ref{Poisson-1}) in the primed frame, the fields 
$\{\delta,\vv,\phi\}$ satisfy the equations of motion (\ref{cont-1})-(\ref{Poisson-1})
in the unprimed frame, with the gravitational potential transforming as
\beq
\phi' = \phi - a^2 (\ddot{\epsilon} + 2 H \dot{\epsilon} ) x^2/2 .
\label{phip-1}
\eeq

This remains valid beyond shell crossing: if the trajectories $\vx'(\vq,t)$ satisfy the
equation of motion (\ref{traj-1}) in the primed frame, the trajectories $\vx(\vq,t)=
(1-\epsilon)\vx'(\vq,t)$ satisfy the equation of motion (\ref{traj-1}) in the unprimed frame,
the gravitational potentials transforming as in Eq.(\ref{phip-1}).

Linearizing over the density contrast, the peculiar velocity, and the perturbation 
$\epsilon$, we have
\beq
\delta_L = \delta_L' + 3\epsilon , \;\;\; \vv_L = \vv_L' - \dot{\epsilon} a \vx , \;\;\;
\phi_L = \phi_L' + a^2 (\ddot{\epsilon}  + 2 H \dot{\epsilon} ) \frac{x^2}{2} .
\label{deltaL-1}
\eeq
In agreement with the remark above, we can again check that if 
$\{\delta_L',\vv_L',\phi_L'\}$ is a valid linear growing mode in the primed frame,
$\{\delta_L,\vv_L,\phi_L\}$ is a valid linear growing mode in the unprimed frame.
Moreover, the density contrast $\delta$ is equal to $\delta'$, up to the dilatation
(\ref{x-d-v-1}), to which is added the uniform contribution $3\epsilon(t)$ that 
corresponds to a homogeneous linear growing mode, as seen from 
Eq.(\ref{eps0-def}).

We can now compute the dependence of small-scale density correlations on
$\epsilon_0$, that is, on changes of the background density.
Thus, we consider the response function
\beq
R^n_{\epsilon_0} = \left \lag \frac{\pl [ \tdelta(\vk_1,t_1) .. \tdelta(\vk_n,t_n) ]}{\pl \epsilon_0} 
\right \rag_{\epsilon_0=0} .
\label{Rn-eps0-def}
\eeq
As described above, adding a nonzero background $\epsilon$ corresponds to
changing the initial background density from the reference $\rhob$ to the primed
density $\rhob'$. This modifies the growth of large-scale structures, as the latter
evolve in a new cosmology, defined by a new set of cosmological parameters.
In particular, starting from a concordance $\Lambda$-CDM flat cosmology with
$\Om+\Omega_{\Lambda}=1$, the change of the background density generates
a curvature term $K/a^2$.
For a given set of initial conditions $\delta_{L0}$, the new field $\delta_{\epsilon_0}$,
measured in the reference frame with the added background $\epsilon$, can be
expressed in terms of the density contrast $\delta'$ in the primed frame, where
$\epsilon$ has been absorbed by the change $\rhob\rightarrow \rhob'$,
through the mapping (\ref{x-d-v-1}). This gives
\beq
\delta_{\epsilon_0}(\vx,t) = (1+3\epsilon) \, \delta'[ (1+\epsilon)\vx,t] + 3 \epsilon ,
\label{map-x-1}
\eeq
which reads in Fourier space as
\beq
\tdelta_{\epsilon_0}(\vk,t) = \tdelta'[(1-\epsilon) \vk,t] + 3 \epsilon\, \delta_D(\vk) .
\label{map-k-1}
\eeq

Next, we use the approximate symmetry described in Sec.~\ref{approx-sym}
to write that the density contrast only depends on the cosmological parameters
through the linear growth rate $D_+(t)$, whence 
$\tdelta'(\vk,t;\Om',\Omega_{\Lambda}',\Omega_K') = \tdelta(\vk,D_+'(t))$,
where $\tdelta(\vk,D_+)$ is the functional that gives the nonlinear density contrast
for any set of cosmological parameters, for a given initial condition of the zero-mean
linear density contrast.
Thus, Eq.(\ref{map-k-1}) writes as
\beq
\tdelta_{\epsilon_0}(\vk,t) = \tdelta[(1-\epsilon) \vk,D_{+\epsilon_0}] 
+ 3 \epsilon \, \delta_D(\vk)  ,
\label{map-k-2}
\eeq
where $D_{+\epsilon_0}$ is the linear growth rate that is modified with respect to the
initial $D_+$ by the perturbation $\epsilon$.
Then, the derivative of the density contrast with respect to $\epsilon_0$ reads as
\beq
\left . \frac{\pl \tdelta(\vk,t)}{\pl\epsilon_0} \right |_{\epsilon_0=0} = 
\left . \frac{\pl D_{+\epsilon_0}}{\pl\epsilon_0} \right |_0 \frac{\pl \tdelta}{\pl D_+} 
- D_+(t) \vk \cdot \frac{\pl\tdelta}{\pl\vk} ,
\label{d-delta-d-eps0-1}
\eeq
where we disregarded the Dirac factor that does not contribute for wave numbers
$\vk\neq 0$.

We need to compute the dependence of the linear growing mode $D_{+\epsilon_0}$
on $\epsilon_0$. The linear growth rates $D_+$ and $D_+'$ obey 
Eq.(\ref{Dlin-1}), with unprimed and primed Hubble and density factors.
Writing $D_+'(t)=D_+(t) + y(t)$, where $y$ is of order $\epsilon$, we obtain at
linear order
\beq
\ddot{y} + 2 H \dot{y} - 4\pi\cG \rhob y = 2 \dot{D}_+ \dot{\epsilon}
+ 12\pi\cG \rhob D_+ \epsilon .
\label{y-t-1}
\eeq
By definition of the matter density cosmological parameter $\Om$, the mean density
also obeys
\beq
4\pi\cG\rhob = \frac{3}{2} \Om H^2 \simeq \frac{3}{2} \frac{\dot{D}_+^2}{D_+^2} ,
\label{rho-Dp-1}
\eeq
where in the last expression we again used the approximation $\Om \simeq f^2$
associated with the approximated symmetry discussed in 
Sec.~\ref{approx-sym}.
Then, using $\eta=\ln D_+$ as the time coordinate, Eq.(\ref{y-t-1}) becomes
\beq
\frac{\dd^2 y}{\dd\eta^2} + \frac{1}{2} \frac{\dd y}{\dd\eta} - \frac{3}{2} y 
= \frac{13}{2} \, \epsilon_0 \, e^{2\eta} ,
\eeq
which gives
\beq
y(t) = \frac{13}{7} \, \epsilon_0 \, D_+(t)^2 , \;\;\;\;
\left . \frac{\pl D_{+\epsilon_0}}{\pl\epsilon_0} \right |_0 = \frac{13}{7} D_+(t)^2 .
\label{y-t-2}
\eeq
This result was also obtained in App.D of \cite{Baldauf2011}.
Then, Eq.(\ref{d-delta-d-eps0-1}) also writes as
\beq
\left . \frac{\pl \tdelta(\vk,t)}{\pl\epsilon_0} \right |_{\epsilon_0=0} = 
D_+(t) \left[ \frac{13}{7}  \frac{\pl \tdelta}{\pl \ln D_+} 
- \vk \cdot \frac{\pl\tdelta}{\pl\vk} \right] ,
\label{d-tdelta-d-eps0-2}
\eeq
which corresponds in configuration space to
\beq
\left . \frac{\pl \delta(\vx,t)}{\pl\epsilon_0} \right |_{\epsilon_0=0} = 
D_+(t) \left[ 3 \, \delta + \frac{13}{7}  \frac{\pl \delta}{\pl \ln D_+} 
+ \vx \cdot \frac{\pl\delta}{\pl\vx} \right] .
\label{d-delta-d-eps0-2}
\eeq
Eq.(\ref{d-delta-d-eps0-2}) also follows from Eq.(\ref{map-x-1}), where we
disregard the constant factor $3\epsilon$ because we consider small-scale
wave numbers $k\neq 0$. In configuration space, this means that these relations
are valid up to a constant density, which is irrelevant because we consider small-scale
structures and disregard zero-mode (infinitely large-scale) normalizations.

This gives the impact of a large-scale uniform density perturbation, or a change of the
background density, on the small-scale nonlinear density field.
Indeed, from Eq.(\ref{deltaL-1}), the variable $\epsilon_0$ corresponds to 
a change of the linear density contrast of
\beq
\Delta \delta_{L0} = 3 \epsilon_0.
\label{d-deltaL0-eps0}
\eeq

\subsection{Consistency relations}
\label{Equal-time}

\subsubsection{One large-scale wave number}

The comparison of Eq.(\ref{d-deltaL0-eps0}) with Eq.(\ref{eps-x-1}) gives
\beq
\epsilon_0 = \frac{\varepsilon}{3} \int \dd\vx' W(x') C_{L0}(x') ,
\label{eps0-eps-1}
\eeq
where we used the fact that $W$ is a large-scale window and the integral 
over $\vx'$ is independent of the position $\vx$ in the small-scale region, at leading 
order in the ratio of scales.
This gives
\beq
C^n_W =  \frac{1}{3} \int \dd\vx' W(x') C_{L0}(x') 
\frac{\pl \lag \delta(\vx_1,t_1) .. \delta(\vx_n,t_n) \rag_{\epsilon_0}}{\pl\epsilon_0} ,
\label{Cn-x-eps0-1}
\eeq
and using Eq.(\ref{d-delta-d-eps0-2}), we obtain
\beqa
C^n_W & \rightarrow & \int \dd\vx' W(x') C_{L0}(x') \sum_{i=1}^n \frac{D_{+i}}{3} 
\left( 3 + \frac{13}{7}  \frac{\pl}{\pl \ln D_{+i}} \right. \nonumber \\
&& \left. + \vx_i \cdot \frac{\pl}{\pl\vx_i} \right) 
\lag \delta(\vx_1,t_1) .. \delta(\vx_n,t_n) \rag .
\label{Cn-x-eps0-2}
\eeqa
The small-scale correlation $\lag \delta(\vx_1,t_1) .. \delta(\vx_n,t_n) \rag$ is
invariant through translations, thanks to statistical homogeneity.
However, the dilatation operators $\vx_i\cdot\pl/\pl\vx_i$ break this
invariance when the times $t_i$ are not identical.
Indeed, as described in Sec.~\ref{large-scale}, the change of the background density
$\rhob$ due to the uniform density fluctuation (\ref{d-deltaL0-eps0}) leads
to a modified Hubble flow. This breaks the translation invariance for 
different-time statistics, as defining a different Hubble flow selects the origin
from which comoving particles separate along with the global expansion.
This is due to the large-scale approximation for the filter $W$ where we
considered that the small-scale region has a zero-width at the center of the
larger-scale fluctuation.
To explicitly enforce this configuration, we set the center of the modified Hubble flow
at the center of the small-scale region by writing
\beqa
C^n_W \! & = & \! \int \!  \dd\vx' W(x') C_{L0}(x') \sum_{i=1}^n \frac{D_{+i}}{3} 
\biggl [ 3 + \frac{13}{7}  \frac{\pl}{\pl \ln D_{+i}} \;\;\;  \nonumber \\
&& \hspace{-0.5cm}  + \biggl ( \vx_i - \frac{1}{n} \sum_{j=1}^n \vx_j \biggl ) \cdot 
\frac{\pl}{\pl\vx_i} \biggl ] \lag \delta(\vx_1,t_1) .. \delta(\vx_n,t_n) \rag ,
\label{Cn-x-eps0-3}
\eeqa
which is explicitly invariant through uniform translations of $\{\vx_i\}$, as all terms
only depend on relative distances.
In agreement with the remark above, Eq.(\ref{Cn-x-eps0-3}) is identical to
Eq.(\ref{Cn-x-eps0-2}) when all times are equal, $D_{+1}=..=D_{+n}$.
Using the definition (\ref{Cn-Wx-1}), this gives the configuration-space
consistency relation
\beqa
\!\!\! \int \!\! \dd\vx' W(x') \, \lag \delta_{L0}(\vx') \delta(\vx_1,t_1) .. \delta(\vx_n,t_n) \rag
\!\! & = & \!\!\! \int \!\! \dd\vx' \, W \, C_{L0}(x') \nonumber \\
&& \hspace{-5.8cm} \times \sum_{i=1}^n \frac{D_{+i}}{3} 
\biggl [ 3 + \frac{13}{7}  \frac{\pl}{\pl \ln D_{+i}} + \biggl ( \vx_i - \frac{1}{n} 
\sum_{j=1}^n \vx_j \biggl ) \cdot \frac{\pl}{\pl\vx_i} \biggl ] 
\nonumber \\
&& \hspace{-5.8cm} \times \; \lag \delta(\vx_1,t_1) .. \delta(\vx_n,t_n) \rag .
\label{Cn-x-eps0-4}
\eeqa
As explained above, this relation holds in the large-scale limit for the filter
$W$, and up to uniform offsets for the density contrasts $\delta_i$
[i.e., the equality is valid when one integrates both sides with arbitrary weights
$W_i(\vx_i)$ that have a zero mean, $\int\dd\vx_i W_i(\vx_i) =0$].

It is often more convenient to work in Fourier space (because the linearized
equations of motion become diagonal).
Because of statistical homogeneity, multispectra contain a Dirac factor that we
explicitly factor out by defining
\beq
\lag \tdelta(\vk_1) .. \tdelta(\vk_n) \rag =
\lag \tdelta(\vk_1) .. \tdelta(\vk_n) \rag'  \; \delta_D(\vk_1 \!+ .. +\! \vk_n) .
\label{multi-spectra}
\eeq
To simplify the analysis, it is convenient to consider $\lag \tdelta_1 .. \tdelta_n \rag'$
in Eq.(\ref{multi-spectra}) as a function of $\{\vk_1,..,\vk_{n-1}\}$ only,
by substituting for $\vk_n=-(\vk_1+..+\vk_{n-1})$.
Using the invariance through translations of 
$\lag \delta_1 .. \delta_n \rag$, which gives 
$\sum_{i=1}^n \pl/\pl\vx_i \cdot \lag \delta_1 .. \delta_n \rag = 0$, 
we can write the dilatation factors of Eq.(\ref{Cn-x-eps0-4}) (denoted
as the overall operator $\cbD$, without the factor $1/3$)
as
\beqa
\cbD \cdot \lag \delta_1 .. \delta_n \rag & = & \biggl \lbrace \sum_{i=1}^{n-1} 
\biggl[ D_{+i} \, (\vx_i \!-\! \vx_n) + \frac{D_{+n} \!-\! D_{+i}}{n} \nonumber \\
&& \times \sum_{j=1}^{n-1} (\vx_j \!-\! \vx_n) 
\biggl] \cdot \frac{\pl}{\pl\vx_i} \biggl \rbrace \; \lag \delta_1 .. \delta_n \rag .
\eeqa
Using the Fourier transform of the density correlation as in Eq.(\ref{multi-spectra})
and integrating over $\vk_n$, this yields
\beqa
\cbD \cdot \lag \delta_1 .. \delta_n \rag & = & \int \! \dd\vk_1 .. \dd\vk_{n-1} \, 
\biggl \lbrace \biggl [ \sum_{i=1}^{n-1} D_{+i} \, \vk_i \cdot \frac{\pl}{\pl\vk_i} 
\nonumber \\
&& \hspace{-1.5cm} + \sum_{i,j=1}^{n-1} \frac{D_{+n}-D_{+i}}{n} \, 
\vk_i \cdot \frac{\pl}{\vk_j} \biggl ]
e^{\ii \sum_{i=1}^{n-1} \vk_i \cdot (\vx_i-\vx_n)} \biggl \rbrace \nonumber \\
&& \hspace{-1.5cm} \times \; \lag \tdelta_1 .. \tdelta_n \rag' .
\eeqa
Integrating by parts, using $\pl/\pl\vk_n \cdot \lag \tdelta_1 .. \tdelta_n \rag'=0$ and
$\vk_1+..+\vk_n=0$, this also writes as
\beqa
\cbD \cdot \lag \delta_1 .. \delta_n \rag \! & = & \! - \int \! \dd\vk_1 .. \dd\vk_{n-1} \, 
e^{\ii \sum_{i=1}^{n-1} \vk_i \cdot (\vx_i-\vx_n)} \biggl [ 3 \frac{n-1}{n} 
\nonumber \\
&& \hspace{-2cm} \times \sum_{i=1}^n D_{+i} + \sum_{i,j=1}^n ( \delta^K_{i,j} - 
\frac{1}{n} ) D_{+i} \, \vk_i \cdot \frac{\pl}{\pl\vk_j} \biggl ] 
\lag \tdelta_1 .. \tdelta_n \rag' ,
\eeqa
where $\delta^K_{i,j}$ is the Kronecker symbol.
Therefore, Eq.(\ref{Cn-x-eps0-4}) reads in Fourier space as
\beqa
\int \frac{\dd\vOm_{\vk'}}{4\pi} \lag \tdelta_{L0}(\vk') \tdelta(\vk_1,t_1) .. 
\tdelta(\vk_n,t_n) \rag'_{k'\rightarrow 0} =  P_{L0}(k') \nonumber \\
&& \hspace{-7.8cm} \times \sum_{i=1}^n D_{+i} \biggl [ \frac{1}{n} + 
\frac{13}{21} \frac{\pl}{\pl\ln D_{+i}} - \sum_{j=1}^n (\delta^K_{i,j} - \frac{1}{n} ) 
\frac{\vk_i}{3} \! \cdot \! \frac{\pl}{\pl\vk_j} \biggl ]
\nonumber \\
&& \hspace{-7.8cm} \times \; \lag \tdelta(\vk_1,t_1) .. \tdelta(\vk_n,t_n) \rag' ,
\label{tCn0-1}
\eeqa
where $\vOm_{\vk'}$ is the unit vector along the direction of $\vk'$.
On large scales we recover the linear theory, with $\tdelta(\vk',t') \simeq
D_+(t') \tdelta_{L0}(\vk')$. Thus, Eq.(\ref{tCn0-1}) also writes as
\beqa
\int \frac{\dd\vOm_{\vk'}}{4\pi} \lag \tdelta(\vk',t') \tdelta(\vk_1,t_1) .. 
\tdelta(\vk_n,t_n) \rag'_{k'\rightarrow 0} =  P_L(k',t') \nonumber \\
&& \hspace{-8.2cm} \times \sum_{i=1}^n \frac{D_{+i}}{D_+(t')} \biggl [ \frac{1}{n} + 
\frac{13}{21} \frac{\pl}{\pl\ln D_{+i}} - \sum_{j=1}^n (\delta^K_{i,j} - \frac{1}{n} ) 
\frac{\vk_i}{3} \! \cdot \! \frac{\pl}{\pl\vk_j} \biggl ]
\nonumber \\
&& \hspace{-8.2cm} \times \; \lag \tdelta(\vk_1,t_1) .. \tdelta(\vk_n,t_n) \rag' .
\label{tCn-1}
\eeqa
Because we wrote the operator that acts over $\lag \tdelta_1 .. \tdelta_n\rag'$
in a symmetric form, in Eqs.(\ref{tCn0-1})-(\ref{tCn-1}) we can use any
appropriate form for $\lag \tdelta_1 .. \tdelta_n \rag'$ [i.e., we can write
the $n-$point correlation as a function of 
$\{\vk_1,..,\vk_{i-1},\vk_{i+1},..,\vk_n\}$, as $\vk_i$ can be replaced by
$-(\vk_1+..+\vk_{i-1}+\vk_{i+1}+..+\vk_n)$ for any index $i$, 
or keep it as a function of the $n$ wavenumbers $\{\vk_1,..,\vk_n\}$,
because of the constraint $\vk_1+..+\vk_n=0$].

In contrast with the kinematic consistency relations that express the transport
of small-scale structures by large-scale fluctuations 
\cite{Kehagias2013,Peloso2013,Creminelli2013,Kehagias2013a,Peloso2013a,Creminelli2013a,Valageas2013b},
the angular averaged relations (\ref{tCn-1}) do not vanish when all times are
equal. Indeed, they go beyond this kinematic effect and express the deformation
of small-scale structures by a large-scale isotropic curvature of the gravitational
potential.
When all times are equal, $t'=t_1=..=t_n=t$, Eq.(\ref{tCn-1})
becomes
\beqa
\int \frac{\dd\vOm_{\vk'}}{4\pi} \lag \tdelta(\vk',t) \tdelta(\vk_1,t) .. 
\tdelta(\vk_n,t) \rag'_{k'\rightarrow 0} =  P_L(k',t) \;\;\; \nonumber \\
&& \hspace{-7cm} \times \biggl [ 1 + \frac{13}{21} \frac{\pl}{\pl\ln D_+} 
- \frac{1}{3} \sum_{i=1}^n \frac{\pl}{\pl\ln k_i} \biggl ]
\nonumber \\
&& \hspace{-7cm} \times \; \lag \tdelta(\vk_1,t) .. \tdelta(\vk_n,t) \rag' ,
\label{tCn-2}
\eeqa
where we used $\vk_1+..+\vk_n=0$.

A nice property of the single-time consistency relation (\ref{tCn-2}) is that
it only involves single-time correlations on both sides
(as opposed for instance to a relation that would involve the partial time derivative with
respect to only one time $t_i$ in the right hand side).
Moreover, thanks to the approximate symmetry discussed in 
Sec.~\ref{approx-sym}, both sides involve density correlations in the same (our)
universe. This is because, although the effect of a large-scale curvature is
similar to a change of cosmological parameters, the approximate symmetry
allows us to express the density correlations in the modified cosmology in terms
of the correlations measured in the original universe, through a rescaling of
space and time coordinates.
Therefore, the angular averaged consistency relations (\ref{tCn-2})
can be measured and tested in our Universe. However, this requires measuring
the evolution with time of the density correlations to estimate the time derivative
in the right hand side.

\subsubsection{Bispectrum}

In practice, one does not measure density correlations up to very high orders,
which become increasingly noisy, and most observational constraints
from density correlations come from the 2-point and 3-point correlations.
This corresponds in Fourier space to the power spectrum 
$P(k_1;t_1,t_2)= \lag \tdelta(\vk_1,t_1) \tdelta(\vk_2,t_2)\rag'$  and bispectrum
$B(k_1,k_2,k_3;t_1,t_2,t_3) = \lag \tdelta(\vk_1,t_1) \tdelta(\vk_2,t_2) 
\tdelta(\vk_3,t_3) \rag'$.
(In practice, one measures single-time statistics, but for completeness we
also consider the different-time statistics.)
Taking into account the constraint $\vk'+\vk_1+\vk_2=0$ by writing
$\vk_1 = \vk - \vk'/2$ and $\vk_2 = -\vk-\vk'/2$, with some arbitrary wave number
$\vk$, Eq.(\ref{tCn-1}) reads for $n=2$ as
\beqa
\int \frac{\dd\vOm_{\vk'}}{4\pi} \; B \!\left( \vk', \vk-\frac{\vk'}{2},-\vk-\frac{\vk'}{2}
;t',t_1,t_2 \right)_{k'\rightarrow 0} =  \;\;\; \nonumber \\
&& \hspace{-7.5cm} \frac{P_L(k',t')}{D_+(t')} \biggl [ \frac{D_{+1}+D_{+2}}{2} 
\biggl ( 1 - \frac{1}{3} \frac{\pl}{\pl\ln k} \biggl ) \nonumber \\
&& \hspace{-7.5cm} + \frac{13}{21} \biggl ( D_{+1}^2 \frac{\pl}{\pl D_{+1}}
+ D_{+2}^2 \frac{\pl}{\pl D_{+2}} \biggl ) \biggl ] P(k;t_1,t_2) .
\label{B-1}
\eeqa
When all times are equal to $t$, this becomes, in agreement with
Eq.(\ref{tCn-2}),
\beqa
\int \frac{\dd\vOm_{\vk'}}{4\pi} \; B \!\left( \vk', \vk-\frac{\vk'}{2},-\vk-\frac{\vk'}{2}
;t \right)_{k'\rightarrow 0} =  P_L(k',t) \;\;\; \nonumber \\
&& \hspace{-7cm} \times \biggl [ 1 + \frac{13}{21} \frac{\pl}{\pl\ln D_+} 
- \frac{1}{3} \frac{\pl}{\pl\ln k} \biggl ] P(k,t) .
\label{B-2}
\eeqa

\subsubsection{Several large-scale wave numbers}
\label{several}

It is possible to generalize the single-time consistency relation (\ref{tCn-2}) to $\ell$
large-scale wave numbers by an iterative procedure, as long as they follow a
hierarchy $k_1' \ll k_2' \ll .. \ll k_{\ell}'$, because the angular average and the
derivative $\pl/\pl\ln k$ commute.
This gives
\beqa
k_{j}' \ll k_{j+1}' : \;\;\; \int \prod_{j=1}^{\ell} \frac{\dd\vOm_{\vk_j'}}{4\pi} \;
\lag \prod_{j=1}^{\ell} \tdelta(\vk_j') \prod_{i=1}^n \tdelta(\vk_i) \rag'_{k'_j\rightarrow 0} 
= \nonumber \\
&& \hspace{-5cm} \cL_1' .. \cL_{\ell}' \cdot \lag \tdelta(\vk_1) .. \tdelta(\vk_n) \rag' ,
\label{tCln-1}
\eeqa
where the operators $\cL_j'$ read as
\beqa
\cL_j' & = & P_L(k_j',t) \biggl [ 1+ \frac{13}{21} \frac{\pl}{\pl\ln D_+} 
- \frac{1}{3} \sum_{m=j+1}^{\ell} \frac{\pl}{\pl\ln k_m'} \;\;\;
 \nonumber \\
&&  - \frac{1}{3} \sum_{i=1}^n \frac{\pl}{\pl\ln k_i} \biggl ] .
\eeqa
The operators $\cL_j'$ do not commute. This comes from the fact that
the relation (\ref{tCln-1}) is obtained from the iterated use of 
Eq.(\ref{tCn-2}), where the large-scale limits are taken in a specific order,
starting with $k_1'$ and finishing with $k_{\ell}'$, in agreement with the
hierarchy $k_1' \ll k_2' \ll .. \ll k_{\ell}'$.
Expanding the product $\cL_1' .. \cL_{\ell}'$ gives increasingly long expressions
as the number $\ell$ of large-scale modes grows and we do not pursue this matter here.

\subsubsection{Multi-component case}

If there are several fluids in the system, for instance when we separate
dark matter and baryons, each component $(\alpha)$ follows the equations
of motion (\ref{cont-1})-(\ref{Euler-1}) or (\ref{cont-2})-(\ref{Euler-2}),
written in terms of each doublet $\{\delta^{(\alpha)},\vv^{(\alpha)}\}$ or
$\{\delta^{(\alpha)},\vu^{(\alpha)}\}$. 
The gravitational potential $\varphi$ now obeys the Poisson equation
$\nabla^2\varphi = (3\Om)/(2f^2) \sum_{\alpha} (\Omega_{(\alpha)}/\Om) 
\delta^{(\alpha)}$.
Because all fluids are subject to the same gravitational potential,
in the linear growing mode all density contrasts $\delta^{(\alpha)}$ and
velocities $\vv^{(\alpha)}$ are equal (and the mean density
ratios $\Omega_{(\alpha)}/\Om$ remain constant). Then, assuming as for the
single-fluid case that decaying modes have had time to become negligible,
the single-fluid consistency relations (\ref{tCn-1})-(\ref{tCln-1}) extend
at once to the multi-fluid case.

This is no longer true when some fluids are subject to non-gravitational
forces that introduce new scales. For instance, when pressure forces or
astrophysical processes, such as outflows from supernovae, have an
impact on the baryon dynamics, the invariance with respect to changes of
cosmological parameters (in the approximation $\Om/f^2 \simeq 1$)
is broken. Indeed, these new processes generically introduce different
explicit dependences on density and time scales that cannot be reduced to
functions of $\Om/f^2$.
Thus, our results only hold in the regime where gravity is the dominant force.

\subsection{Comments on the derivation of the consistency relations}
\label{Analysis}

The consistency relations 
(\ref{tCn-1})-(\ref{tCn-2}), and their derivation through the effect of large-scale
perturbations in Sec.~\ref{large-scale}, may appear somewhat counterintuitive,
and we comment in this section on some points that may seem puzzling.

As explained in Sec.~\ref{Corr-resp}, the consistency relations derive from the
exact relation (\ref{Cn-Rn-1}), which shows how the correlation between $n$ 
small-scale nonlinear modes and a large-scale linear mode can be obtained
from the mean response of the small-scale modes to the large-scale mode.
Then, the procedure presented in Secs.~\ref{large-scale} and \ref{Equal-time}
describes how to obtain this response by identifying the large-scale mode with
a change of the background density.

This gives rise to Eqs.(\ref{map-x-1})-(\ref{map-k-1}), which are the first nontrivial
result (no approximation has been used at this stage).
However, they may already raise several questions that we address in turns.

First, it might seem that Eqs.(\ref{map-x-1})-(\ref{map-k-1}), which give the impact of
a large-scale perturbation onto the small-scale nonlinear density contrast, contradict
linear theory where different Fourier modes are known to decouple.
This is not the case, because in the linear regime these equations read as
$\delta_{\epsilon_0} = \delta ' + 3 \epsilon$  for Eq.(\ref{map-x-1}),
as in the first Eq.(\ref{deltaL-1}), and as 
$\tilde{\delta}_{\epsilon_0} = \tilde{\delta}' + 3 \epsilon \delta_D(\vk)$.
This merely means that in the linear regime different growing mode solutions 
superpose.
Since $\epsilon_0$ corresponds to a uniform density shift, it gives rise to the
Dirac factor in Eq.(\ref{map-k-1}), and we recover the fact that in the linear regime 
different wave numbers are decoupled (hence a perturbation at $\vk' \rightarrow 0$
only affects the same mode with $\vk=\vk'\rightarrow 0$).
Therefore, there is no contradiction between Eqs.(\ref{map-x-1})-(\ref{map-k-1})
and linear theory.

In the nonlinear regime, different wave numbers become coupled, in agreement
with the rescaling and amplification found in the first terms in the right-hand-sides
in Eqs.(\ref{map-x-1})-(\ref{map-k-1}).
However, one may worry that a first-order expansion over $\epsilon$ may not be sufficient
to evaluate the response of small-scale nonlinear density contrasts.
For instance, let us consider a single one-dimensional plane wave for the
small-scale initial perturbation. This fluctuation will collapse at some time $t_*$ to
build an infinite-density 2D sheet, of zero thickness. If we add a small linear large-scale
mode $\epsilon$, the collapse time $t_*$ will be changed to a slightly different
value $t_*+\Delta t_*$. Then, for any nonzero $\epsilon$ and $\Delta t_*$, the impact on
the densities at time $t_*$ along the position of the pancake is not small and not
proportional to $\epsilon$ (since it is the difference between infinity and a finite value).
However, this is a pathological case that appears with a zero probability: in this example 
it only occurs at the precise time $t_*$ at the position of the pancake.
At later times, there remains an infinite density sheet in both the zero-$\epsilon$
and nonzero-$\epsilon$ cases, and the main change is to slightly modify its position and mass.
This still gives a large deviation if we measure the density contrast at the precise position
of the sheet, but this is no longer the case if we perform some smoothing over an arbitrarily 
small window or if we consider the Fourier-space density contrast $\tilde\delta(\vk)$
(which actually integrates over all space).

Here, it is more convenient to think in terms of the particle trajectories $\vx(\vq,t)$.
Then, using the fact that the Fourier-space density contrast can be written as a function
of the particle positions, in a form similar to the 1D expression (\ref{tdelta-xq}) used below,
we can see that as long as particle trajectories display a first-order expansion over
$\epsilon$, a similar expansion holds for the Fourier-space density contrast, even though
shell crossings and infinite-density sheets may have appeared.
This point will become obvious in Sec.~\ref{check-1D}, where we show that the
consistency relations are actually exact in 1D until shell crossing, and even remain
exact after shell crossings if we consider the Zel'dovich dynamics itself instead of the
1D gravitational dynamics (both systems only being identical before shell crossing).

In 3D, even more pathological examples can be found, where particle trajectories
themselves are singular with respect to small perturbations. 
For instance, spherically symmetric solutions with purely radial trajectories
are strongly unstable with respect to non-spherical perturbations and the linear growth rate
actually diverges \cite{Valageas2002b}. Then, infinitesimal perturbations are sufficient
to initiate the virialization of the cloud and generate significant transverse motions.
However, these cases again appear with a zero probability (initial conditions are not exactly
spherically symmetric with purely radial motions) and they should not impair the derivation
presented in Secs.~\ref{large-scale} and \ref{Equal-time}.

The identification of the response of the system to a large-scale mode with the response to
a change of the background density might seem puzzling, as even a large-scale mode
is constrained to have a zero mean. However, it is rather clear that from the point of view
of a small-scale region located at the center of the larger-scale perturbation, the latter acts
as a uniform change of the background density.
This is explicitly shown by the fact that Eq.(\ref{eps-t}), which describes the evolution of 
the difference between two close bakgrounds, is identical to the evolution equation 
(\ref{Dlin-1}) of the linear growth rates $D_{\pm}(t)$.  
This is because they follow from the same fundamental set of equations, that
describe the gravitational force. 

Explicit examples that also clarify these points are presented in the next 
Sec.~\ref{checks}, where we check the consistency relations at lowest order of perturbation
theory for the bispectrum in 3D, and at all orders for all polyspectra in 1D.

\section{Explicit checks}
\label{checks}

The angular averaged consistency relations (\ref{tCn-1})-(\ref{tCn-2}) 
are valid at all orders of perturbation theory and also beyond the
perturbative regime, including shell crossing effects, within the accuracy
of the approximation $\Om/f^2 \simeq 1$ (and as long as gravity is the
dominant process).

We now provide two explicit checks of the angular averaged consistency relations 
(\ref{tCn-1})-(\ref{tCn-2}).
First, we check these relations for the lowest-order case $n=2$, that is, for the
bispectrum, at lowest order of perturbation theory.
Second, we present a fully nonlinear and nonperturbative check, for arbitrary
$n-$point polyspectra, in the one-dimensional case.

\subsection{Perturbative check}
\label{perturbative}

Here we briefly check the consistency relations for the
lowest order case, $n=2$, given by Eqs.(\ref{B-1})-(\ref{B-2}), at lowest order of
perturbation theory.
At this order, the density bispectrum reads as \cite{Bernardeau2002}
\beqa
B(\vk_1,\vk_2,\vk_3;t_1,t_2,t_3) & = & D_{+1} D_{+2} D_{+3}^2 P_{L0}(k_1) 
P_{L0}(k_2) \;\;\; \nonumber \\
&& \hspace{-3cm} \times \left[ \frac{10}{7} + \left( \frac{k_1}{k_2}
+\frac{k_2}{k_1} \right)  \frac{\vk_1\cdot\vk_2}{k_1 k_2}
+ \frac{4}{7} \left( \frac{\vk_1\cdot\vk_2}{k_1 k_2} \right)^2 \right] \nonumber \\
&& \hspace{-3cm} + \; 2 \mbox{ perm.}
\label{bispec-1}
\eeqa
where ``2 perm.'' stands for two other terms that are obtained from permutations
over the indices $\{1,2,3\}$.
In the small $k'$ limit we obtain
\beqa
B(\vk',\vk_1,\vk_2;t',t_1,t_2)_{k'\rightarrow 0} \! & = & \! D_+' D_{+1} D_{+2}^2 
P_{L0}(k') P_{L0}(k_1) \nonumber \\
&& \hspace{-3.5cm} \times \left[ \frac{10}{7} + \frac{\vk_1\cdot\vk'}{k'^2} 
+ \frac{4}{7} \left( \frac{\vk_1\cdot\vk'}{k_1 k'} 
\right)^2 \right] + \; ( 1 \leftrightarrow 2 ) .
\label{bispec-2}
\eeqa
Here we used the fact that the term in the bracket in Eq.(\ref{bispec-1}) vanishes
as $k_3^2$ for $k_3\rightarrow 0$, whereas $P_{L0}(k_3) \sim k_3^{n_s}$ with
$n_s \lesssim 1$. 
[If this is not the case, that is, there is very little initial power 
on large scales, we must go back to the consistency relation in the form 
of Eq.(\ref{tCn0-1}) rather than Eq.(\ref{tCn-1}). However, this is not necessary
in realistic models.]
Then, taking into account the constraint $\vk'+\vk_1+\vk_2=0$ by writing
$\vk_1 = \vk - \vk'/2$ and $\vk_2 = -\vk-\vk'/2$ as in Eq.(\ref{B-1}), 
and expanding Eq.(\ref{bispec-2}) over $k'$, we obtain
\beqa
B_{k'\rightarrow 0} \! & = & \! D_+' D_{+1} D_{+2}^2 P_{L0}(k') \biggl [ P_{L0}(k) 
\left( \frac{13 \!+\! 8\mu^2}{14} + \frac{k}{k'}\mu \right)  \nonumber \\
&& - \frac{\dd P_{L0}(k)}{\dd\ln k} \frac{\mu^2}{2} \biggl ] + \; ( 1 \leftrightarrow 2 ) ,
\label{bispec-3}
\eeqa
where $\mu=(\vk\cdot\vk')/(k k')$.
The integration over angles gives
\beqa
\int \frac{\dd\vOm_{\vk'}}{4\pi} B_{k'\rightarrow 0} & = & D_+' D_{+1} D_{+2} 
\frac{D_{+1}+D_{+2}}{2} P_{L0}(k') \nonumber \\
&& \times \left[ \frac{47}{21} P_{L0}(k) - \frac{1}{3} 
\frac{\dd P_{L0}(k)}{\dd\ln k} \right] .
\label{bispec-4}
\eeqa

On the other hand, the right hand side of Eq.(\ref{B-1}) writes as
\beqa
\int \frac{\dd\vOm_{\vk'}}{4\pi} B_{k'\rightarrow 0} \!\! & = & D_+' P_{L0}(k') 
\biggl [ \frac{D_{+1}+D_{+2}}{2} \biggl ( 1 - \frac{1}{3} \frac{\pl}{\pl\ln k} \biggl ) 
\nonumber \\
&& \hspace{-2.3cm} + \frac{13}{21} \biggl ( D_{+1}^2 \frac{\pl}{\pl D_{+1}}
+ D_{+2}^2 \frac{\pl}{\pl D_{+2}} \biggl ) \biggl ] D_{+1} D_{+2} P_{L0}(k) 
\eeqa
and we recover Eq.(\ref{bispec-4}).
This provides a check of  Eq.(\ref{B-1}) and also of the single-time relation (\ref{B-2}), 
which is a particular case of Eq.(\ref{B-1}). 
Alternatively, the same procedure applied to the single-time bispectrum provides
a direct check of Eq.(\ref{B-2}).

Therefore, we have checked the angular average consistency relation
(\ref{tCn-1}) for the bispectrum, at leading order of perturbation theory,
within the approximate symmetry $\Om/f^2 \simeq 1$ discussed in
Sec.~\ref{approx-sym}.
In this explicit check, the use of this approximate symmetry appears at the level of
the expression (\ref{bispec-1}) of the bispectrum, which only involves the linear 
growing mode $D_+$.
An exact calculation would give prefactors for the terms in the bracket
that show new but weak dependences on time and cosmology (and are unity for
the Einstein-de Sitter case) \cite{Bernardeau2002}.
These deviations from Eq.(\ref{bispec-1}) are usually neglected 
[for instance, when the cosmological constant is zero, they were shown to be well 
approximated by factors like $(\Om^{-2/63}-1)$ that are very small over the range
of interest \cite{Bouchet1992}].

\subsection{1D nonlinear check}
\label{check-1D}

The explicit check presented in Sec.~\ref{perturbative} only applies up to the
lowest order of perturbation theory. Because the goal of the consistency relations 
is precisely to go beyond low-order perturbation theory, it is useful to
obtain a fully nonlinear check. This is possible in one dimension, where the
Zel'dovich solution \cite{ZelDovich1970} becomes exact (before shell crossing)
and all quantities can be explicitly computed.
Because of the change of dimensionality, we also need to rederive the 1D form of
the consistency relations. We present the details of our computations in
App.~\ref{sec-1D-example}, and only give the main steps in this section.

\subsubsection{1D Equations of motion}

First, as described in App.~\ref{1D-motion}, using the change of variables
(\ref{eta-1D}) the 1D equations of motion can be written as
\beq
\frac{\pl\delta}{\pl \eta} + \frac{\pl}{\pl x} [ (1+\delta) u ] = 0 ,  \label{cont1D-3}
\eeq
\beq
\frac{\pl u}{\pl \eta} + [ \kappa(t) - 1 ]  u + u \frac{\pl u}{\pl x} = 
- \frac{\pl \varphi}{\pl x} ,  
\label{Euler1D-3}
\eeq
\beq
\frac{\pl^2\varphi}{\pl x^2} = \kappa(t) \delta ,
\label{Poisson1D-3}
\eeq
where we introduced the factor $\kappa(t)$ defined by
\beq
\kappa(t) = 4\pi \cG(t) \bar{\rho}(t) \frac{D_+(t)^2}{\dot{D}_+(t)^2} .
\label{kappa-def}
\eeq
As explained in App.~\ref{1D-motion}, we generalized the system to the case
of a time-dependent Newton's constant $\cG(t)$. This allows us to obtain
scale factors $a(t)$ that expand forever as power laws, as in Eq.(\ref{at-1D}),
in a fashion that mimics the Einstein-de Sitter 3D cosmology.

Thus, $\kappa(t)$ plays the role of the ratio $3\Omega_{\rm m}/(2f^2)$ encountered in
the 3D case in Eqs.(\ref{cont-2})-(\ref{Poisson-2}).
In the 1D cosmology (\ref{at-1D}) with $c_1=0$, it is a constant given by
$\kappa_0 = - (\alpha+2)/(\alpha+1)$.
Then, the 3D approximation $\Om/f^2 \simeq 1$ used in the main text corresponds
in our 1D toy model to the approximation $\kappa \simeq \kappa_0$. That is, we
neglect the dependence of $\kappa$ on the cosmological parameters (here the
coefficient $c_1$) and the dependence on the background is fully contained in the
change of variables (\ref{eta-1D}). This is the 1D approximate symmetry that is the
equivalent of the 3D approximate symmetry used in the previous sections.
The generalization to the case of a time-dependent Newton's constant is not
important at a formal level, because it does not modify the form of the equations of
motion. However, it is necessary for this approximate symmetry to make practical sense,
so that we can find a regime where $\kappa$ is approximately constant [here,
around $c_1=0$ with the choice (\ref{alpha-def})].

The fluid equations (\ref{cont1D-3})-(\ref{Poisson1D-3}) only apply to the 
single-stream regime, but we can again go beyond shell crossings by using the
equation of motion of trajectories, which reads as
\beq
\frac{\pl^2 x}{\pl\eta^2} +  \left[ \kappa(t)  - 1 \right] 
\frac{\pl x}{\pl\eta} = -  \frac{\pl \varphi}{\pl x} ,
\label{traj1D-2}
\eeq
where $\varphi$ is the rescaled gravitational potential (\ref{Poisson1D-3}).
This is the 1D version of Eq.(\ref{traj-2}) and it explicitly shows that particle
trajectories obey the same approximate symmetry, before and after shell crossings.

\subsubsection{1D consistency relations}
\label{1D-consistency}

To derive the consistency relations, we can follow the method described in the previous 
sections for the 3D case.
In a fashion similar to the analysis of Sec.~\ref{large-scale}, we first derive in
App.~\ref{1D-background-perturbation} the impact of a large-scale linear density
perturbation on the small-scale nonlinear density field. 
This gives in Fourier space
\beq
\left . \frac{\pl \tdelta(k,t)}{\pl\epsilon_0} \right |_{\epsilon_0=0} = 
D_+(t) \left[  \frac{\pl \tdelta}{\pl \ln D_+} 
- k \frac{\pl\tdelta}{\pl k} \right] ,
\label{d-tdelta-d-eps0-1D-2}
\eeq
which corresponds in configuration space to
\beq
\left . \frac{\pl \delta(x,t)}{\pl\epsilon_0} \right |_{\epsilon_0=0} = 
D_+(t) \left[ \delta +  \frac{\pl \delta}{\pl \ln D_+} 
+ x \frac{\pl\delta}{\pl x} \right] .
\label{d-delta-d-eps0-1D-2}
\eeq
As expected, we recover the same forms as in 
Eqs.(\ref{d-tdelta-d-eps0-2})-(\ref{d-delta-d-eps0-2}), but with different
numerical coefficients because the dimension of space has changed.

To obtain the 1D consistency relations, we follow the method described in 
Sec.~\ref{Equal-time}. The factors $1/3$ are replaced by unity and we use the
result (\ref{d-tdelta-d-eps0-1D-2}). 
Then, Eq.(\ref{tCn-1}) becomes
\beqa
\frac{1}{2} \sum_{\pm k'} \lag \tdelta(k',t') \tdelta(k_1,t_1) .. 
\tdelta(k_n,t_n) \rag'_{k'\rightarrow 0} =  P_L(k',t') \nonumber \\
&& \hspace{-7.2cm} \times \sum_{i=1}^n \frac{D_{+i}}{D_+(t')} \biggl [ \frac{1}{n} + 
\frac{\pl}{\pl\ln D_{+i}} - \sum_{j=1}^n (\delta^K_{i,j} - \frac{1}{n} ) 
k_i \frac{\pl}{\pl k_j} \biggl ]
\nonumber \\
&& \hspace{-7.2cm} \times \; \lag \tdelta(k_1,t_1) .. \tdelta(k_n,t_n) \rag' .
\label{tCn-1D-1}
\eeqa
The 3D angular average $\int \dd\vOm_{\vk'}/(4\pi)$ of Eq.(\ref{tCn-1}) is replaced
by the 1D average $\frac{1}{2} \sum_{\pm k'}$ over the two directions of $k'$
(i.e., the two signs of $k'$).
When all times are equal, $t'=t_1=..=t_n=t$, Eq.(\ref{tCn-2})
becomes
\beqa
\hspace{-1cm} \frac{1}{2} \sum_{\pm k'} \lag \tdelta(k',t) \tdelta(k_1,t) .. 
\tdelta(k_n,t) \rag'_{k'\rightarrow 0} =  P_L(k',t) \nonumber \\
&& \hspace{-7.3cm} \times  \biggl [ 1 + 
\frac{\pl}{\pl\ln D_+} - \sum_{i=1}^n \frac{\pl}{\pl \ln k_i} \biggl ]
 \; \lag \tdelta(k_1,t) .. \tdelta(k_n,t) \rag' .
\label{tCn-1D-2}
\eeqa

\subsubsection{1D explicit checks}
\label{1D-explicits}

As is well known, in 1D the Zel'dovich approximation \cite{ZelDovich1970}
is actually exact before shell crossing. We briefly check this property in 
App.~\ref{1D-Zeldovich} on the generalized system (\ref{cont1D-2})-(\ref{Poisson1D-2}), 
where Newton's constant can vary with time.
Then, using the conservation of matter, $(1+\delta) \dd x = \dd q$, the Fourier-space 
nonlinear density contrast can be written as \cite{Taylor1996,Valageas2007a}
\beq
\tdelta(k) = \int \frac{\dd x}{2\pi} e^{-\ii k x} \delta(x) = 
\int \frac{\dd q}{2\pi} e^{-\ii k x(q)} - \delta_D(k) .
\label{tdelta-xq}
\eeq
Using the solution Eq.(\ref{xq-1D}), and disregarding the Dirac term for $k \neq 0$,
this gives the usual expression
\beq
\tdelta(k,t) = \int \frac{\dd q}{2\pi} \, e^{-\ii k q + \int \dd k' e^{\ii k' q} \frac{k}{k'} \tdelta_L(k',t) } .
\label{tdelta-Z-1}
\eeq
The explicit nonlinear expression (\ref{tdelta-Z-1}) allows us to check the
1D consistency relations (\ref{tCn-1D-1})-(\ref{tCn-1D-2}).
We present in App.~\ref{1D-check} two different checks.

First, in App.~\ref{Impact-1D}, following the derivation of the consistency relations, 
we check that the impact of a large-scale perturbation on the nonlinear density contrast
is given by Eq.(\ref{d-tdelta-d-eps0-1D-2}). This is the key relation from which the
consistency relations derive and it provides a first 1D nonlinear check.

Second, in App.~\ref{Explicit-1D}, we directly check the consistency relations 
(\ref{tCn-1D-1})-(\ref{tCn-1D-2}) from the explicit expressions of the density polyspectra,
without going through the intermediate step that considers the impact of a
large-scale linear perturbation on the small-scale nonlinear density contrast.
Thus, this also provides a check of the reasoning that underlies the derivation
of these relations.

\subsubsection{1D nonlinear and nonperturbative validity}

The explicit checks presented in App.~\ref{1D-check} 
show that the consistency relations (\ref{tCn-1D-1})-(\ref{tCn-1D-2}) are actually
exact in the 1D case, before shell crossing.
Here there is a simplification, as compared with the 3D case, that makes the
consistency relations exact beyond the approximation of constant $\kappa$.
This is because the nonlinear density contrast (\ref{tdelta-Z-1}) only depends on the
linear growing mode $D_+(t)$ and the approximation of constant $\kappa$ is not 
needed.
This is also apparent in Eqs.(\ref{cont1D-4})-(\ref{Poisson1D-4}) 
or Eq.(\ref{xt-1D-4}).
For the solution (\ref{xq-1D}), the right-hand sides in Eqs.(\ref{Euler1D-4}) 
and (\ref{xt-1D-4}) cancel out and we obtain a dynamics that only involves the
linear growing mode $D_+(t)$ as the rescaled time coordinate.
Then, for this class of solutions the symmetry that is the basis of the consistency
relations is actually exact (before shell crossing).

This 1D toy model provides an explicit nonlinear check of the consistency relations.
This also ensures that they are valid to all orders of perturbation theory. 
However, because the single-stream equations of motion 
(\ref{cont1D-2})-(\ref{Poisson1D-2}) and the solution (\ref{tdelta-Z-1}) 
only apply before shell crossing, which is a nonperturbative effect, this 1D
gravitational toy model does not provide an explicit check in terms of
nonperturbative shell crossing contributions.

On the other hand, we can also consider the Zel'dovich solution (\ref{xq-1D}), and 
the associated density contrast (\ref{tdelta-Z-1}), as a second 1D toy model
(which is no longer related to ``gravitational'' forces).
This second system only coincides with the 1D gravitational dynamics before
shell crossing and departs from it afterwards, but it is also a well defined system at all
times. Then, we can apply the same reasoning that underlies the derivation of the
consistency relations to this system, which satisfies the same symmetries.
From this point of view, the solution (\ref{tdelta-Z-1}) now provides an
explicit nonlinear check of the consistency relations that also applies beyond
shell crossing.

As noticed above, it happens that in this 1D case the function $\kappa(t)$ does not
appear in the solution (\ref{tdelta-Z-1}), so that the approximation 
$\kappa \simeq \kappa_0$ is not needed and the consistency relations
(and the underlying symmetry) are exact, up to shell crossing for the 1D gravitational
dynamics, and even beyond shell crossing when we consider the second toy model
defined by the solution (\ref{xq-1D}).
Therefore, the time dependence of the scale factor $a(t)$ and of $\kappa(t)$
is irrelevant and we could as well keep the usual case of a time-independent
Newton's constant $\cG$, with a finite collapse time of the system.
Nevertheless, it is nice to consider the generalized 1D system (\ref{alpha-def})
that can mimic the usual 3D cosmological expansion up to infinite time.
Moreover, if we consider the true 1D gravitational dynamics, which departs
from the Zel'dovich solution (\ref{xq-1D}) after shell crossing, the function
$\kappa(t)$ will appear in the nonperturbative shell crossing terms, when the
right-hand side in Eq.(\ref{xt-1D-4}) no longer cancels out.
Then, the consistency relations are only approximate in the shell crossing regime,
up to the accuracy of the $\kappa \simeq \kappa_0$ approximation.

\section{Conclusion}
\label{conclusion}

As explained in the previous sections, the angular averaged consistency relations
(\ref{tCn-1})-(\ref{tCn-2}) only rely on the approximate symmetry
discussed in Sec.~\ref{approx-sym}, which states that the dependence on
cosmological parameters can be absorbed through the mapping
$t\rightarrow D_+$ of the time coordinate (within the approximation
$\Om/f^2 \simeq 1$).
Therefore, our results are not restricted to the perturbative regime
and also apply to small nonlinear scales governed by shell-crossing effects,
as long as the approximation $\Om/f^2 \simeq 1$ is sufficiently accurate.

We have also pointed out that these relations are actually exact in 1D gravitational systems
until shell crossing, and even beyond shell crossing for the Zel'dovich dynamics
itself (which departs from 1D gravity after shell crossing).

It is difficult to extend these results to the galaxy number density field 
in a rigorous fashion, because galaxy formation is not expected to satisfy
the approximate symmetry of Sec.~\ref{approx-sym}. For instance,
cooling processes and star formation introduce new time and density scales,
which means that the time-coordinate mapping $t\rightarrow D_+$ is not sufficient to 
absorb all cosmological dependence.

However, these effects are likely to be subdominant if we assume, as in halo models,
that galaxies are closely related to the dark matter density field.
Thus, in most analytical
approaches, one writes the galaxy number density fluctuations as a functional
of the matter density fluctuations at the same time, 
$\delta_{\rm g}(\vx,t) = \delta_{\rm g}[\delta(.,t)]$.
In the popular local bias model, this is simplified as a function of the local density 
contrast, smoothed over some scale $R$,
$\delta_{\rm g}(\vx,t) = \delta_{\rm g}[\delta_R(\vx,t)] 
= \sum_{n=1}^{\infty} \frac{b_n}{n!} \delta_R(\vx,t)^n$.
However, this introduces an explicit model dependence, especially as the
transformation (\ref{x-d-v-1}) modifies the amplitude of the density contrast,
which leads to a change of the number of halos in the framework of halo models.
Then, the simplest method to derive consistency relations for the galaxy
distribution from the matter density relations (\ref{tCn-1})-(\ref{tCln-1})
is to start from an explicit galaxy bias model, that allows one to express
galaxy correlations in terms of matter correlations. Then, one can directly
use Eqs.(\ref{tCn-1}) and (\ref{tCln-1}) and obtain constraints on the
galaxy correlations.
The simplest case is the constant linear bias model, where $\delta_{\rm g}=
b_1 \delta$, with a scale-independent bias $b_1$ that only depends on the galaxy
type (mass, luminosity,..). Then, Eqs.(\ref{tCn-1}) and (\ref{tCln-1}) directly
extend to the galaxy density field, up to a factor $b_1$.

In more general models of galaxy clustering, one takes into account
higher orders or keeps a functional dependence, $\delta_{\rm g}(\vx) = 
\delta_{\rm g}[\delta(.)]$. In particular, one can use some derivative expansion
of this functional, so that the galaxy density field also involves the deformation
tensor or higher-order derivatives of the smoothed density field
(e.g., \cite{Desjacques2013}). 
Then, the general relation (\ref{Cn-Rn-1}), where the nonlinear matter density
contrasts $\tdelta(\vk_i,t_i)$ are replaced by the galaxy number density contrasts
$\tdelta_{\rm g}(\vk,t_i)$, is still valid, because it only assumes that the field
$\tdelta_{\rm g}$ is a functional of the Gaussian field $\tdelta_{L0}$
\cite{Valageas2013b}. This allows one to write again Eqs.(\ref{Cn-Wx-2})
or (\ref{Cn-Wk-2}) for the galaxy density field.
Next, one can write the derivatives $\cD\delta_{\rm g}/\cD\delta_{L0}$ in terms
of derivatives of the matter density contrast $\delta$, and use 
Eqs.(\ref{d-tdelta-d-eps0-2}) or (\ref{d-delta-d-eps0-2}). However, in the generic
case this gives expressions that involve the matter density field and cannot
be written back in terms of the galaxy density field in a simple manner.

A detection (beyond the range authorized by the finite accuracy of the 
approximation $\Om/f^2 \simeq 1$) of a violation of Eqs.(\ref{tCn-1}) and 
(\ref{tCln-1}), written in terms of the galaxy number density if that is possible,
would signal a breakdown of the underlying galaxy biasing scheme.
To remove this degeneracy, one can also use weak gravitational lensing
observations, which directly probe the matter density field.

On the other hand, these consistency relations rely on the Gaussian
hypothesis for the initial conditions. Indeed, this assumption is used to derive
Eq.(\ref{Cn-Rn-1}) \cite{Valageas2013b}, which is the basis of subsequent
relations. 
Then, a violation of the angular averaged consistency relations
could signal primordial non-Gaussianities.
Other possible interpretations could be effects from nonzero decaying
modes, or a departure from the ``standard'' cosmological
scenarios, for instance a modified-gravity model where the approximation 
$\Om/f^2 \simeq 1$ is strongly violated or where new terms in the equations of
motion show an explicit dependence on cosmology (on the background density)
that has not the form $\Om/f^2$.

Apart from these observational aspects, these angular averaged consistency
relations (\ref{tCn-1})-(\ref{tCln-1}) might be used as a check of numerical
simulations or algorithms.
From a more theoretical perspective, they could also help designing models for 
the matter  density correlations.
In particular, perturbative approaches, which often explicitly use the
approximation $\Om/f^2 \simeq 1$ to simplify the analysis, attempt to go beyond
the standard perturbation theory by including partial resummations of higher-order
diagrams. They can also be seen as different closure schemes, where one 
implements different truncations of the infinite hierarchy between $n$- and
$n+1$-point density correlations (the standard perturbation theory amounts
to set all correlations above some finite order $N$ to zero, while resummation schemes
can be seen as assuming a specific ansatz for the $N+1$ correlation, expressed
in terms of the lower-order ones, to close the hierarchy at order $N$).
Then, Eqs.(\ref{tCn-1}) and (\ref{tCln-1}) might serve as a guideline
to write this $N+1$ correlation in terms of the lower-order correlations.
We leave such investigations to future works.

\begin{acknowledgments}

We thank F. Vernizzi for discussions.
This work is supported in part by the French Agence Nationale de la Recherche
under Grant ANR-12-BS05-0002.

\end{acknowledgments}

\appendix

\section{1D example}
\label{sec-1D-example}

It is interesting to check the consistency relations obtained in this paper on
a simple one-dimensional example that can be exactly solved.
This is provided by the Zel'dovich dynamics \cite{ZelDovich1970}, which is exact in 1D 
(before shell crossing).

\subsection{1D equations of motion}
\label{1D-motion}

In physical coordinates, the 1D continuity, Euler and Poisson equations, read as
\beq
\frac{\pl\rho}{\pl t} + \frac{\pl}{\pl r} (\rho \nu) = 0 ,  \label{cont1D-1}
\eeq
\beq
\frac{\pl\nu}{\pl t} + \nu \frac{\pl \nu}{\pl r} = - \frac{\pl \Psi}{\pl r} ,  
\label{Euler1D-1}
\eeq
\beq
\frac{\pl^2\Psi}{\pl r^2} = 4 \pi \cG(t) \rho ,  
\label{Poisson1D-1}
\eeq
where $\rho(r,t)$ and $\nu(r,t)$ are the 1D density and velocity fields. Here we 
generalized the 1D gravitational dynamics to the case of a time-dependent 
Newton's constant $\cG(t)$.
The background solution corresponds to $\nu = H r$ and 
$\Psi= 2\pi \cG \bar{\rho} r^2$, 
with $H=\dot{a}/a$. The 1D continuity equation yields $\bar\rho = \bar{\rho}_0 /a$
and the Euler equation gives
\beq
\frac{\ddot{a}}{a} = - 4\pi \cG(t) \bar{\rho} ,
\label{Friedman1D-1}
\eeq
which corresponds to the Friedman equation. In the 3D case, the expansion of the
universe gives $\bar\rho \propto a^{-3}$, which dilutes the gravitational attraction 
(as $1/r^2$), and there are ever-expanding solutions (without cosmological constant). 
In the 1D case, we have $\bar\rho \propto a^{-1}$ and the gravitational force is not
diluted by the expansion (as is well known, in 1D the gravitational force is constant
and independent of the distance between particles before shell crossing).
Then, there are no ever-expanding solutions and the system collapses after a finite
time. However, by generalizing to a time-dependent Newton's constant $\cG(t)$, we
can again obtain solutions that expand forever. Thus, we can consider the power-law
models
\beq
-2 < \alpha < -1 : \;\;\; \cG(t) = \cG_0 \left( \frac{t}{t_0} \right)^{\alpha} ,
\label{alpha-def}
\eeq
which lead to the expansion laws
\beq
a(t) = - \frac{4\pi \cG_0 \bar{\rho}_0 t_0^2}{(\alpha+1)(\alpha+2)} 
\left( \frac{t}{t_0} \right)^{\alpha+2} + c_1 t ,
\label{at-1D}
\eeq
where $c_1$ is an arbitrary integration constant. The case $c_1=0$ is the 1D version
of the standard 3D Einstein-de Sitter cosmology, while the term $c_1 t$ plays the
role of the 3D curvature term.
 
We can now switch to comoving coordinates, with $x=r/a$, $v=\nu-H r$, 
$\rho=\bar{\rho}(1+\delta)$, $\phi=\Psi+a \ddot{a} x^2/2$, and we obtain the 1D
version of Eqs.(\ref{cont-1})-(\ref{Poisson-1}),
\beq
\frac{\pl\delta}{\pl t} + \frac{1}{a} \frac{\pl}{\pl x} [ (1+\delta) v ] = 0 ,  
\label{cont1D-2}
\eeq
\beq
\frac{\pl v}{\pl t} + H v + \frac{1}{a} v \frac{\pl v}{\pl x} = - \frac{1}{a} 
\frac{\pl \phi}{\pl x} ,  
\label{Euler1D-2}
\eeq
\beq
\frac{\pl^2\phi}{\pl x^2} = 4 \pi \cG(t) \bar\rho a^2 \delta .  
\label{Poisson1D-2}
\eeq
Linearizing these equations, we obtain the evolution equation of the linear modes of
the density contrast. It takes the same form as the usual 3D equation (\ref{Dlin-1}),
\beq
\ddot{D} + 2 H(t) \dot{D} - 4\pi\cG(t) \rhob(t) D = 0 ,
\label{Dlin1D-1}
\eeq
but with a time-dependent Newton's constant and the 1D scale factor (\ref{at-1D}).
In particular, in the case where $c_1=0$ in Eq.(\ref{at-1D}), which corresponds to the
3D Einstein-de Sitter cosmology, with a scale factor $a(t)$ that keeps expanding
forever but at a decelerated rate, we have the power-law linear growing and decaying
modes
\beq
c_1 =0 : \;\;\; D_+(t) \propto t^{-\alpha-1} , \;\;\; D_-(t) \propto t^{-\alpha-2} .
\label{Dlin1D-t}
\eeq

In a fashion similar to the change of variables (\ref{eta-3D}), we make the change of
variables
\beq
\eta = \ln D_+ , \;\;\; v = \frac{a \dot{D}_+}{D_+} u , 
\;\;\; \phi = \left( \frac{a \dot{D}_+}{D_+} \right)^2 \varphi ,
\label{eta-1D}
\eeq
and we obtain the equations of motion (\ref{cont1D-3})-(\ref{kappa-def}) given
in the main text.
We can again go beyond shell crossings by using the
equation of motion of trajectories (\ref{traj1D-2}).

\subsection{1D background density perturbation}
\label{1D-background-perturbation}

To derive the 1D consistency relations, we can follow the method described in the main text 
for the 3D case and first consider the impact of small changes to the background density.
As in Eq.(\ref{eps-def}), we consider two universes with close cosmological
parameters,
\beq
a' = a [1-\epsilon(t) ] , \;\;\;  \rhob' = \rhob [ 1+\epsilon(t) ] ,
\label{eps1D-def}
\eeq
and substituting into the ``Friedmann equation'' (\ref{Friedman1D-1}) we obtain
\beq
\ddot{\epsilon} + 2 H \, \dot{\epsilon} - 4\pi\cG(t) \rhob \, \epsilon =0 .
\label{eps1D-t}
\eeq
Again, we recover the evolution equation (\ref{Dlin1D-1}) of the linear density modes 
and we can write $\epsilon(t) = \epsilon_0 D_+(t)$.
Next, the change of frame described in Eq.(\ref{x-d-v-1}) becomes
\beq
x' = (1+\epsilon) x , \;\;\; 
\delta' = \delta - \epsilon (1+\delta) , \;\;\;
v' = v + \dot{\epsilon} a x , 
\label{x-d-v-1-1D}
\eeq
and $\delta_L = \delta_L' + \epsilon$.
This means that, as in Eq.(\ref{map-x-1}), the background density perturbation
$\epsilon$ is absorbed by the change of frame as
\beq
\delta_{\epsilon_0}(x,t) = (1+\epsilon) \, \delta'[ (1+\epsilon) x,t] + \epsilon ,
\label{map-x-1-1D}
\eeq
which reads in Fourier space as
\beq
\tdelta_{\epsilon_0}(k,t) = \tdelta'[(1-\epsilon) k,t] + \epsilon \, \delta_D(k) .
\label{map-k-1-1D}
\eeq
Using the approximate symmetry $\kappa \simeq \kappa_0$, we neglect the
dependence of the dynamics on variations of $\kappa$, so that the impact of the
background only comes through the mapping (\ref{eta-1D}). Therefore, as in
Eq.(\ref{map-k-2}), we write 
$\tdelta_{\epsilon_0}(k,t) = \tdelta[(1-\epsilon) k,D_{+\epsilon_0}] + \epsilon \, \delta_D(k)$,
where $D_{+\epsilon_0}$ is the linear growth rate that is modified with respect to the
initial $D_+$ by the perturbation $\epsilon$.
Then, the derivative of the density contrast with respect to $\epsilon_0$ reads as
\beq
\left . \frac{\pl \tdelta(k,t)}{\pl\epsilon_0} \right |_{\epsilon_0=0} = 
\left . \frac{\pl D_{+\epsilon_0}}{\pl\epsilon_0} \right |_0 \frac{\pl \tdelta}{\pl D_+} 
- D_+(t) k \frac{\pl\tdelta}{\pl k} ,
\label{d-delta-d-eps0-1D-1}
\eeq
where we disregarded the Dirac factor that does not contribute for wave numbers
$k\neq 0$.

Next, writing again the linear growing mode in the primed frame as
$D_+'(t) = D_+(t)+y(t)$, and substituting into Eq.(\ref{Dlin1D-1}) with the primed
background, we obtain the 1D version of Eq.(\ref{y-t-1}) as
\beq
\ddot{y} + 2 H \dot{y} - 4\pi\cG \rhob y = 2 \dot{D}_+ \dot{\epsilon}
+ 4\pi\cG \rhob D_+ \epsilon .
\label{y-t-1D-1}
\eeq
Using Eqs.(\ref{kappa-def}) and (\ref{Dlin1D-1}), this can be written in terms of the
time coordinate $\eta=\ln D_+$ as
\beq
\frac{\dd^2 y}{\dd\eta^2} + (\kappa-1) \, \frac{\dd y}{\dd\eta} - \kappa \, y 
= ( \kappa+2) \, \epsilon_0 \, e^{2\eta} ,
\eeq
which gives (using the approximation of constant $\kappa$)
\beq
y(t) = \epsilon_0 \, D_+(t)^2 , \;\;\;\;
\left . \frac{\pl D_{+\epsilon_0}}{\pl\epsilon_0} \right |_0 = D_+(t)^2 .
\label{y1D-t-2}
\eeq
Then, Eq.(\ref{d-delta-d-eps0-1D-1}) also writes as
Eq.(\ref{d-tdelta-d-eps0-1D-2}) in Fourier space, and Eq.(\ref{d-delta-d-eps0-1D-2})
in configuration space,
where again we disregarded the constant factor $\epsilon$ because we consider
small-scale wave numbers with $k \neq 0$.

Next, this gives the 1D consistency relations (\ref{tCn-1D-1})-(\ref{tCn-1D-2})
as described in Sec.~\ref{1D-consistency}.

\subsection{Zel'dovich solution}
\label{1D-Zeldovich}

Making the change of variable \cite{Vergassola1994}
\beq
v = a \dot{D}_+ \, w , \;\;\; \phi = 4 \pi \cG \bar{\rho} a^2 D_+ \, \psi ,
\eeq
and using $D_+(t)$ as the time coordinate, the equations of motion 
(\ref{cont1D-2})-(\ref{Poisson1D-2}) can be written as
\beq
\frac{\pl\delta}{\pl D_+} + \frac{\pl}{\pl x} [ (1+\delta) w ] = 0 ,  \label{cont1D-4}
\eeq
\beq
\frac{\pl w}{\pl D_+} + w \frac{\pl w}{\pl x} = - \frac{\kappa}{D_+} 
\left( \frac{\pl \psi}{\pl x} + w \right)  
\label{Euler1D-4}
\eeq
\beq
\frac{\pl^2\psi}{\pl x^2} = \frac{\delta}{D_+} .  
\label{Poisson1D-4}
\eeq
Then, we can check that $w=-\pl\psi/\pl x$ is a solution of the equations of motion,
with the continuity and Euler equations reducing to $\pl w/\pl D_+ + w \pl w / \pl x = 0$.
This gives the solution
\beq
x(q,t) = q + D_+(t) s_{L0}(q) ,
\label{xq-1D}
\eeq
for the trajectories of the particles, where $q$ is the Lagrangian coordinate and 
$s(q,t) = D_+ s_{L0}$ the nonlinear displacement field, which is identical to the linear
displacement field in this 1D case.
This solution breaks down after shell crossing, as the gravitational force on a particle
changes when there is some exchange of matter between the left and right sides of
this particle. Then, the fluid equations no longer apply and we must solve the
equation of motion of the particles, which reads as
\beq
\frac{\pl^2 x}{\pl D_+^2} = - \frac{\kappa}{D_+} 
\left( \frac{\pl x}{\pl D_+} + \frac{\pl \psi}{\pl x} \right) .
\label{xt-1D-4}
\eeq

The displacement field is related to the linear density contrast by
\beq
\delta_L(q,t) = D_+(t) \delta_{L0}(q) \;\;\; \mbox{with} \;\;\; \delta_{L0} = 
- \frac{\dd s_{L0}}{\dd q} ,
\eeq
which also reads as
\beq
s(q,t) = \int_{-\infty}^{+\infty} \dd k \; e^{\ii k q} \; \frac{\ii}{k} \tdelta_L(k,t) .
\label{sL0-tdL0}
\eeq
Then, using the conservation of matter, the Fourier-space nonlinear density contrast
can be written as Eq.(\ref{tdelta-Z-1}).

\subsection{Check of the 1D consistency relations}
\label{1D-check}

\subsubsection{Impact of a large-scale perturbation on the nonlinear density contrast}
\label{Impact-1D}

To check the validity of the 1D consistency relations from the exact solution
(\ref{tdelta-Z-1}), we simply need the change of the nonlinear density contrast
$\tdelta(k)$ when we make a small perturbation $\Delta\delta_L$ to the initial
conditions on much larger scales.
Let us consider the impact of a small large-scale perturbation $\Delta \delta_L$ to the
initial conditions. Here we also restrict to even perturbations, 
$\Delta\tdelta_L(-k') =  \Delta\tdelta_L(k')$, as the consistency relations studied in this
paper apply to spherically-averaged statistics, which correspond to the $\pm k'$
averages in the 1D relations (\ref{tCn-1D-1})-(\ref{tCn-1D-2}).
Then, expanding Eq.(\ref{tdelta-Z-1}) up to first order over $\Delta \delta_L$, and over
powers of $k'$, we obtain
\beqa
\hspace{-1cm} k' \rightarrow 0 : \;\;\; \Delta\tdelta(k) & = & 
\left[ \int \dd k' \Delta\tdelta_L(k') \right] \nonumber \\
&& \hspace{-1.5cm} \times \int \frac{\dd q}{2\pi} \, e^{-\ii k q + \int \dd k'' e^{\ii k'' q} 
\frac{k}{k''} \tdelta_L(k'') }  (\ii k q) .
\label{Delta-Z-1}
\eeqa
Here the limit $k'\rightarrow 0$ means that we consider a perturbation of the initial
conditions $\Delta\tdelta_L(k')$ that is restricted to low wave numbers $k'<\Lambda$
with a cutoff $\Lambda$ that goes to zero (i.e., that is much smaller than the wave
numbers $k$ and $2\pi/q$ of interest).

On the other hand, from the expression (\ref{tdelta-Z-1}) we obtain at once the exact
result
\beq
\frac{\pl\tdelta}{\pl\ln D_+} - k \frac{\pl\tdelta}{\pl k} =  \int \frac{\dd q}{2\pi} \, 
e^{-\ii k q  + \int \dd k'' e^{\ii k'' q} \frac{k}{k''} \tdelta_L(k'') }  (\ii k q) .
\label{tdelta-Z-2}
\eeq
The comparison with Eq.(\ref{Delta-Z-1}) gives
\beq
k' \rightarrow 0 : \;\; \Delta\tdelta(k) = \left[ \int \dd k' \Delta\tdelta_L(k') 
\right]  \left( \! \frac{\pl\tdelta(k)}{\pl\ln D_+} - k \frac{\pl\tdelta(k)}{\pl k} \! \right) 
\label{Delta-Z-2}
\eeq

The consistency relations (\ref{tCn-1D-1})-(\ref{tCn-1D-2}) only rely on the
expression (\ref{d-tdelta-d-eps0-1D-2}), which also reads (at linear order over
$\epsilon_0$) as
\beq
\Delta\tdelta(k) = \epsilon_0 \, D_+(t) \, 
\left( \! \frac{\pl\tdelta(k)}{\pl\ln D_+} - k \frac{\pl\tdelta(k)}{\pl k} \! \right) .
\label{Delta-Z-3}
\eeq
Since we have $\epsilon_0 = \Delta \delta_L/D_+$, we recover 
Eq.(\ref{Delta-Z-2}).

\subsubsection{Explicit check on the density polyspectra}
\label{Explicit-1D}

Instead of looking for the impact of a large-scale linear perturbation onto the nonlinear
density contrast, as in Sec.~\ref{Impact-1D}, we can directly check the consistency
relations in their forms (\ref{tCn-1D-1}) or (\ref{tCn-1D-2}).
Considering for simplicity the equal-time polyspectra (\ref{tCn-1D-2}), we define
\beqa
E_n(k';k_1,..,k_n;t) & \equiv &
\lag \tdelta_L(k',t) \tdelta(k_1,t) .. \tdelta(k_n,t) \rag \nonumber \\
&& \hspace{-2.3cm} = D_+ \biggl \lag \tdelta_{L0}(k') \int 
\frac{\dd q_1 .. \dd q_n}{(2\pi)^n} \; e^{-\ii \sum_{j=1}^n  k _j  \, q_j} \nonumber \\
&& \hspace{-2cm} \times \; e^{D_+ \int \dd k/k \; \tdelta_{L0}(k) 
\sum_{j=1}^n k_j \,  e^{\ii k q_j} } \biggl \rag ,
\eeqa
where in the last expression we used Eq.(\ref{tdelta-Z-1}).
The Gaussian average over the initial conditions $\tdelta_{L0}$ gives
\beqa
E_n  & = & - \frac{P_L(k')}{k'} \int \frac{\dd q_1 .. \dd q_n}{(2\pi)^n} \; 
\sum_{j=1}^n k_j \, e^{-\ii k' q_j} \nonumber \\
&& \hspace{-0.7cm} \times \; e^{-\ii \sum_{j=1}^n k_j \, q_j} \;
e^{-D_+^2/2 \int  \dd k/k^2 \; P_{L0}(k) 
\left| \sum_{j=1}^n k_j e^{\ii k q_j} \right|^2 } . \nonumber \\
&&
\eeqa
Making the changes of variable $q_1=q_1'+q_n$, .., $q_{n-1}=q_{n-1}'+q_n$,
the argument of the last exponential does not depend on $q_n$.
Then, the integration over $q_n$ yields a Dirac factor $\delta_D(k'+k_1+..+k_n)$,
that we factor out by defining $E_n = E_n' \delta_D(k'+k_1+..+k_n)$,
with a primed notation as in Eq.(\ref{multi-spectra}), and we replace $k_n$
by $-(k'+k_1+..+k_{n-1})$.
Finally, in the limit $k' \rightarrow 0$ we expand the terms $e^{-\ii k' q_j}$ up to first
order over $k'$, and we obtain
\beqa
k' \rightarrow 0 & : & \;\; E_n' = P_L(k') \int \frac{\dd q_1 .. \dd q_{n-1}}{(2\pi)^{n-1}}
\nonumber \\
&& \hspace{-1.1cm} \times \; \biggl [ 1 + \ii \sum_{j=1}^{n-1} k_j \, q_j \biggl ] 
\; e^{-\ii \sum_{j=1}^{n-1} k_j \, q_j } \nonumber \\
&& \hspace{-1.1cm} \times \; e^{-D_+^2/2 \int \dd k/k^2 \; P_{L0}(k) 
\left|  \sum_{j=1}^{n-1} k_j ( e^{\ii k q_j}-1) \right|^2 } . 
\label{En-1D-1}
\eeqa

Proceeding in the same fashion, the $n-$point polyspectra read as
\beqa
\hspace{-0.9cm} P_n \equiv \lag \tdelta(k_1,t) .. \tdelta(k_n,t) \rag'&&  \nonumber \\
&& \hspace{-3.5cm} = \int \frac{\dd q_1 .. \dd q_{n-1}}{(2\pi)^{n-1}}
\; e^{-\ii \sum_{j=1}^{n-1} k_j \, q_j} \nonumber \\
&& \hspace{-3.2cm} \times \; e^{-D_+^2/2 \int \dd k/k^2 \; P_{L0}(k) 
\left|  \sum_{j=1}^{n-1} k_j ( e^{\ii k q_j}-1) \right|^2 } . 
\label{Pn-1D}
\eeqa
Then, we can explicitly check from the comparison with Eq.(\ref{En-1D-1}) that
we have the relation
\beq
k' \rightarrow 0 : \;\; E_n' = P_L(k') \left[ 1 + \frac{\pl}{\pl\ln D_+} - \sum_{j=1}^{n-1}
\frac{\pl}{\pl\ln k_j} \right] P_n ,
\label{En-Pn}
\eeq
and we recover the consistency relation (\ref{tCn-1D-2}) [in Eq.(\ref{En-Pn}) the
right-hand-side does not involve $k_n$ because it has been replaced by
$-(k_1+..+k_{n-1})$ in Eq.(\ref{Pn-1D}), using the Dirac factor $\delta_D(k_1+..+k_n)$].

\bibliography{ref1}   

\end{document}